\documentclass[aps,groupedaddress,superscriptaddress,amsmath,amssymb,twocolumn,pra,longbibliography]{revtex4-1}
\usepackage{bm}
\usepackage[dvips]{graphicx}
\usepackage{wrapfig,subfigure}
\usepackage{stmaryrd}
\usepackage{bbm}
\usepackage{color}
\usepackage[normalem]{ulem}
\usepackage{enumerate}
\usepackage{epstopdf}
\usepackage{braket}
\def\qb{{\bf q}}

\def\rb{{\bf r}}
\def\Rb{{\bf R}}
\def\pb{{\bf p}}
\def\Pb{{\bf P}}
\def\vb{{\bf v}}

\def\Ef{E_{\rm fl}}

\newcommand{\sumprime}[1]{\sum_{#1}{\vphantom{\sum}}^{\!\!\prime}}

\begin{document}
\author{K. Moskovtsev}
\altaffiliation[Pemanent address: ]{SiTime Corporation, 5451 Patrick Henry Drive Santa Clara, CA 95054. USA}
\affiliation{Department of Physics and Astronomy, Michigan State University, East Lansing, MI 48824, USA}
\altaffiliation[Pemanent address: ]{SiTime Corporation, 5451 Patrick Henry Drive Santa Clara, CA 95054. USA}
\author{M. I. Dykman}
\affiliation{Department of Physics and Astronomy, Michigan State University, East Lansing, MI 48824, USA}

\title{Mobility of a spatially modulated  electron liquid on the helium surface}

\date{\today}
\begin{abstract}
We present the results of large-scale numerical simulations of the mobility of a two-dimensional electron liquid on the helium surface in the presence of a one-dimensional periodic potential. Even where the potential is much weaker than the electron-electron interaction, it can strongly change the mobility. The effect depends on the interrelation between the potential period and the mean interelectron distance. It is most pronounced  where the period is close to the period of the Wigner crystal that would form if the liquid were cooled  to a lower temperature. The results suggest, in particular, that the correlation length in the electron liquid can be found by measuring the mobility in a weak periodic potential. The simulations are based on  the microcopic model of the electron scattering by the excitations in helium.
\end{abstract}
\maketitle

\section{Introduction}
\label{sec:Intro}

Electrons floating above the surface of liquid helium form a peculiar two-dimensional (2D) condensed-matter system.  Its major distinctive features are the absence of  a disorder potential  and the strong electron correlations \cite{Andrei1997,Monarkha2004}. 
The absence of disorder makes it the best-known condensed-matter conductor, with the electron momentum relaxation time $\sim 10^{-7}$~s for $T\lesssim 0.3$~K  \cite{Shirahama1995,Collin2002}. The strong correlations lead to Wigner crystallization for low temperatures \cite{Grimes1979,Fisher1979}, with the Wigner crystal having unusual and not yet entirely understood properties \cite{Rees2016,Rees2017}. 
On the higher-temperature side of the transition, the electrons form a liquid, which displays anomalous classical and quantum magnetotransport \cite{Dykman1979a,Dykman1993b,Lea1998} and numerous nontrivial nonequilibrium phenomena, see \cite{Konstantinov2009,Konstantinov2013,Chepelianskii2015} 
and references therein. We emphasize that, even though the electron liquid is nondegenerate, its dynamics is determined entirely by the electron-electron interaction. The interaction is not a perturbation.

Since the electrons float in free space with no leads attached, a major way of studying them is by measuring the response to a low-frequency electric field or to microwaves. The response to a spatially uniform field is not directly affected by the electron-electron interaction, which preserves the total momentum \cite{Kohn1961}. 
However, this interaction modifies, sometimes dramatically, the short-wavelength electron scattering by helium excitations, in particular by the surface capillary waves (ripplons) and phonons. In turn, this changes the electron transport compared to the single-electron transport, as observed in the experiments mentioned above, cf. also \cite{Buntar'1987,Menna1993,*Barabash-Sharpee2000}. 
While the general picture of the many-electron transport on helium is commonly accepted,  we are not aware of a direct observation of correlations in the electron liquid.

\begin{figure}[h]
\centering
\includegraphics[scale=0.3]{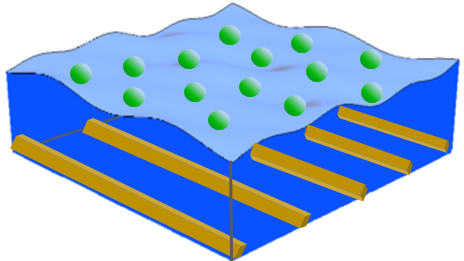} \qquad
\includegraphics[scale=0.1]{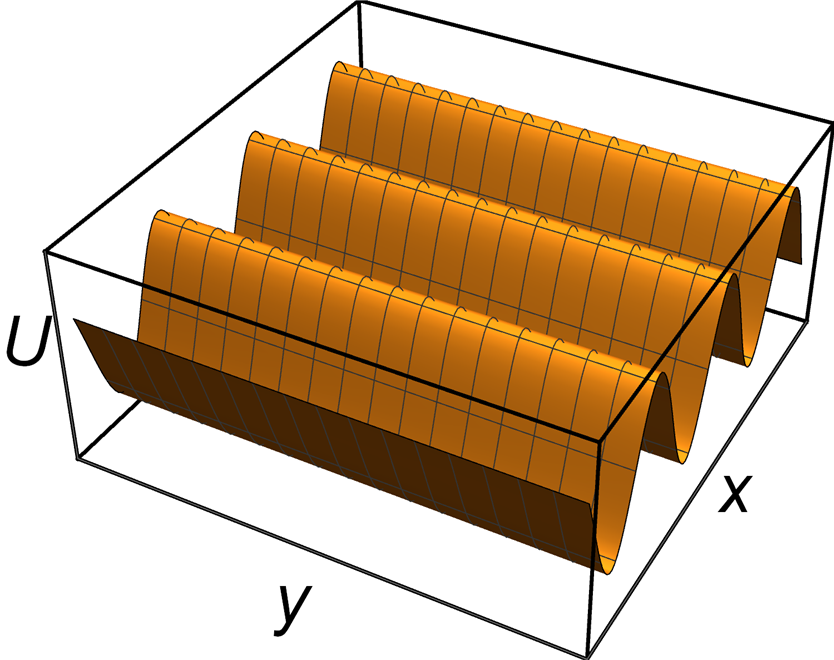}
\caption{Left panel: sketch of the electron liquid floating above the helium surface with a submerged periodic grating of nanowires that create a periodic one-dimensional potential for the electrons. The potential is shown in the right panel. It is essentially sinusoidal, if the electrodes are submerged by a depth that noticeably exceeds the inter-electrode spacing.}
\label{fig:electrodes}
\end{figure}

In this paper we argue that the electron correlations in the liquid can be directly revealed by studying electron transport in the presence of a one-dimensional periodic potential with period close to the mean inter-electron distance. We provide the results of numerical simulations of the electron mobility that substantiate this claim. Since the mean inter-electron distance  on helium is $\sim 1 \mu$m, a potential with the corresponding period can be created by a conventionally grown grating of electrodes submerged beneath the helium surface, as sketched in Fig.~\ref{fig:electrodes}. 

A strong effect of the periodic potential on the electron transport is easy to see already in a single-electron picture. Here, in the classical regime the mobility $\mu_\perp$ transverse to the potential troughs will be thermally activated, $\mu_\perp\propto \exp(-\Delta U/k_BT)$, where $\Delta U$ is the difference between the maximum and the minimum of the potential. Strong correlations in the electron liquid modify this picture in two ways. First, if the potential is much weaker than the electron-electron interaction and has a period very different from the mean distance between the electrons $a_s$, the effect of the potential is partly averaged out, as there are electrons both near the minima and near the maxima of the potential. On the other hand, even where the potential is weak, but its period is close to $a_s$, the electron density can become strongly modulated. In this case, the mobility should strongly depend on the correlation length in the electron liquid, since moving the system with a modulated density as a whole over the periodic potential barriers is impossible, in the limit of a large system. Indeed, our simulations show a strong dependence of $\mu_\perp$ on the correlation length.

Placing a system of interacting particles into a one-dimensional (1D) periodic potential is known to affect the transition from the liquid to the ordered phase, cf. \cite{Wei1998,Radzihovsky2001}. 
The change of the critical temperature and of the very character of the transition is particularly strong in a system with the Coulomb coupling \cite{Kalia1983}. This is because such a system displays a true crystalline order if the period of the potential is equal to the lattice constant of the system \cite{Moskovtsev2019}. Understanding the transport of a  crystal in a periodic potential is interesting and challenging, cf. recent papers \cite{Bylinskii2016,Brazda2018,Zakharov2019a} and references therein. However, for electrons on helium 
 the analysis is complicated by a specific mechanism of scattering by helium excitations \cite{Dykman1997b,Vinen1999}.
Therefore, here we do not study the dynamics of the electron crystal and do not consider the rich area of commensurate-incommensurate transitions in a periodic potential, cf. \cite{Pokrovsky1980,Bak1982,Aubry1983}. Our goal in this paper is to reveal the features of the electron liquid.

Our analysis is based on numerical simulations of a classical 2D electron system on helium. Simulations of 2D electron systems have attracted significant attention over the years, cf.
\cite{Gann1979,Hansen1979,Morf1979,Kalia1981,Kalia1983,Strandburg1988,FangYen1997,Muto1999,Piacente2005,Clark2009,Damasceno2010, Rees2012,Mazars2015}. 
The classical and quantum Monte Carlo as well as molecular dynamics (MD) methods have been employed. However, the MD simulations were done by adding friction forces and uncorrelated noises that drive individual electrons. Such phenomenological description does not describe the dynamics of electrons on helium. Rather, the major mechanism of electron scattering for low temperatures is scattering by surface capillary waves, ripplons \cite{Shikin1974a,Cole1974}. Such scattering is quasielastic, since ripplons are very slow. 

We simulate the mobility directly by taking into account the scattering by ripplons and studying the current induced by a weak electric field applied to the electron system. Such a field unavoidably heats up the electron system. Therefore it is necessary to incorporate a mechanism of energy exchange between the electrons and helium.  An important mechanism of such exchange is scattering by phonons \cite{Dykman2003a,Moskovtsev2019}. Even though the corresponding scattering rate is small compared to the rate of scattering by ripplons, it is sufficient to avoid electron overheating in the range of the fields we study.

The microscopic picture of the electron relaxation for an electron liquid  takes into account the forces that the electrons exert on each other and that affect their scattering by ripplons and phonons \cite{Dykman1979a}; it also relies on the assumption of the fast exchange of energy and momentum between  the electrons \cite{Dykman1979a,Buntar'1987,Dykman1993b}. This exchange is faster than the electron scattering by the helium excitations. The results of the analysis based on this picture are in a good agreement with the experiment, but in fact the assumption has not been tested directly. As we show, numerical simulations suggest a way to carry out this test. However, the central results of the paper refer to the mobility in a periodic potential and its dependence on the amplitude of the potential and the temperature.

Below in Sec.~II we describe the model of the system, briefly outline the simulations and indicate the parameter range where they apply. In Sec. III we present results on the mobility of a uniform electron liquid. In Sec.~IV we describe the mobility in a sinusoidal periodic potential with the period incommensurate with the period of the hexagonal Wigner crystal for the studied electron density; however, we emphasize that our system is not a crystal, the lattice constant of the crystal is used just as the spatial scale. Section ~V is the central part of the paper. It describes the correlations in the electron system and the  dependence of the electron mobility on temperature and the amplitude of the potential where the potential is commensurate with the would-be Wigner crystal. In particular, there is discussed the activated dependence of the mobility on the potential amplitude and its relation to the correlation length in the electron liquid. Section~VI briefly presents the results on the dependence of the electron mobility on the commensurability parameter. Section VII contains concluding remarks.

\section{Many-electron system on helium}
\label{sec:model}

\subsection{The Hamiltonian}

The Hamiltonian of the electron system coupled to the helium excitations has the form 
\begin{align}
\label{eq:Hamiltonian_complete}
H=H_{ee} + H_U + H_{\mathrm He} + H_i.
\end{align}
The term $H_{ee}$ is a sum of the electron kinetic energy and the energy of the electron-electron interaction, whereas $H_U$ is the electron energy in the extrenal potential,
\begin{align}
\label{eq:Hamiltonian_ee}
&H_{ee} = \sum_n\frac{\pb_n^2}{2m_e}+ \frac{1}{2}\sum_{n\neq m}\frac{e^2}{|\rb_n-\rb_m|}, \nonumber\\
&H_U = \sum_n U(\rb_n), \qquad U(\rb) =- A\cos Qx,
\end{align}
Here, $\rb_n = (x_n,y_n)$ and $\pb_n = (p_{xn},p_{yn})$ are the 2D coordinate and momentum of an $n$th electron, and $U(\rb)$ is the external periodic potential. The electrodes creating the potential, see Fig.~\ref{fig:electrodes}, partly screen the electron-electron interaction. The screening depends on the specific geometry, and for reasonably deeply submerged thin electrodes is comparatively weak. The partial screening does not destroy the long-range nature of the electron-electron interaction. Therefore it should not qualitatively change the mobility of the electron liquid. In what follows it is disregarded. We note that, in the experiment, the effect of the screening can be independently tested by varying the height of the helium layer and simultaneously changing the electrode potential. If the distance to the electrodes significantly exceeds the inter-electrode spacing, the potential $U(\rb_n)$ is sinusoidal, as indicated in Eq.~(\ref{eq:Hamiltonian_ee}), with $2\pi/Q$ being the inter-electrode spacing and $A$ being the amplitude of the potential.

The term $H_{\rm He}$ in Eq.~(\ref{eq:Hamiltonian_complete}) is the Hamiltonian of the excitations in the liquid helium that are coupled to the electrons. These are vibrational modes, i.e., ripplons and phonons,  
\begin{align}
\label{eq:He_modes}
\hat H_{\mathrm He} =\sum_{\qb,\,\alpha}  \hbar\omega_{\qb\,\alpha}\hat a^\dagger_{\qb\,\alpha} \hat a_{\qb\,\alpha}. 
\end{align}
Here $\qb$ is the 2D wave vector of a mode, $\omega_{\qb\,\alpha}$ is the mode frequency, and $a^\dagger_{\qb\,\alpha}$ and  $\hat a_{\qb\,\alpha}$ are creation and annihilation operators.  For phonons, the quantum number $\alpha$ is the wave number $q_z>0$ of motion transverse to the surface. Ripplons are surface waves, in this case $\alpha$ is just the label of the vibrational branch.

The Hamiltonian of the electron coupling to the vibrational modes is 
\begin{align}
\label{eq:coupling}
\hat H_i = \sum_n\sum_{\qb,\alpha}V_{\qb\,\alpha}e^{i\qb\hat\rb_n}(\hat a_{\qb\,\alpha}+\hat a_{-\qb\,\alpha}^\dagger). 
\end{align}
The coupling parameters $V_{\qb\,\alpha}$ are well-known. For completeness, they are given in Appendix \ref{sec:simulations}. 
As mentioned previously, electron scattering by ripplons is the major mechanism of the electron momentum relaxation, whereas the scattering by phonons is a major mechanism of the energy relaxation; the other energy relaxation mechanism is two-ripplon scattering. Since we allow for energy relaxation primarily to avoid electron heating when the electrons are additionally driven by an electric field, we consider only one energy relaxation mechanism, the phonon scattering.

To study the conductivity of the electron liquid, we add to the Hamiltonian the term
\begin{align}
\label{eq:drive_term}
H_{\rm d}  = eE_{\rm d}\sum_nx_n,
\end{align}
where $E_{\rm d}$ is the driving field. This field is weak, we chose it to be much smaller than the fluctuational field that drives an electron due to electron density fluctuations and is $\sim n^{3/4}(k_BT)^{1/2}$ \cite{Dykman1979a,FangYen1997}, where $n_s$ is the electron density. As we show, for the values of $E_d$ we use, the electron heating is weak.

The major parameters that characterize the electron liquid are the short-wavelength plasma  frequency $\omega_p$ and the plasma parameter $\Gamma$, which is the ratio of the typical interaction energy per electron $E_C$ to the kinetic energy (the analog of $r_s$ in degenerate systems),
\begin{align}
\label{eq:parameters}
\omega_p = &(2\pi e^2 n_s^{3/2}/m_e)^{1/2}, \qquad \Gamma = E_C/k_BT, \nonumber\\
&E_C=e^2(\pi n_s)^{1/2}.
\end{align}
Parameter $\omega_p$ gives the typical rate of the momentum and energy exchange between the electrons, whereas $\Gamma$ shows how strong the electron correlations are. In the absence of the periodic potential, Monte Carlo simulations suggest that the classical electron liquid crystallizes into a Wigner crystal for $\Gamma\approx 140$, cf. \cite{Mazars2015} and references therein.

\begin{figure}[h]
\includegraphics[scale=0.5]{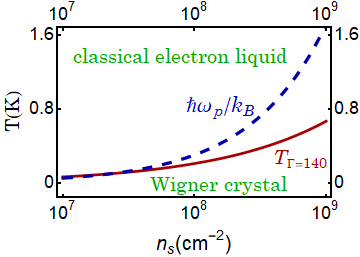}  
	\caption{The characteristic boundaries of the classical motion (blue dashed line) and Wigner crystallization (red solid line) on the temperature/electron density plane.  The plasma frequency $\omega_p=(2\pi e^2 n_s^{3/2}m)^{1/2}$ is the characteristic frequency of electron vibrations about their quasi-equilibrium position in a liquid or solid with the electron density $n_s$. The temperature $T_\Gamma$ is the temperature given by the relation $T= e^2(\pi n_s)^{1/2}/k_B\Gamma$, where $\Gamma$ is the plasma parameter and $n_s$ is the electron density. Recent Monte-Carlo simulations of the classical system \cite{Mazars2015} set the value of $\Gamma$ for crystallization in the classical limit at $\Gamma=140$.}
\label{fig:crystal_boundaries}
\end{figure}

We note that sometimes,  particularly in comparing the simulated temperature of the Wigner crystallization with the experiment, the conditions  that the electron system is  classical and nondegenerate are  tacitly presumed to be equivalent. However, a nondegenerate electron system is not necessarily classical, see Fig.~\ref{fig:crystal_boundaries}. In the absence of a transverse magnetic field, the characteristic temperature that separates the classical and quantum electron dynamics can be chosen as $\hbar\omega_p/k_B$, where $\omega_p$ is given by Eq.~(\ref{eq:parameters}). This temperature  is an analog of the Debye temperature of the Wigner crystal, but it also characterizes the electron dynamics in the liquid phase. It is shown by the dashed line in Fig.~\ref{fig:crystal_boundaries}. The majority of experiments on Wigner crystallization on helium were done for the electron densities $n_s>10^8{\rm cm}^{-2}$, where the simulated transition temperature is essentially in the quantum regime. A comparison with classical simulations requires going to lower densities or higher temperatures. Our simulations refer to the temperatures $T>\hbar\omega_p/k_B$.

\subsection{Simulations}
\label{subsec:SImulations}

We simulate the electron dynamics in a periodic potential and a superimposed uniform electric field by integrating the many-electron equations of motion. In contrast to the standard molecular dynamics simulations, to find the electron mobility and to describe the features of the many-electron dynamics on helium, we explicitly incorporate scattering by the helium excitations into the equations of motion. This is done by considering scattering as a random event in which the electron momentum and energy change.  The probability of scattering of a given electron within a short time interval is determined by the instantaneous value of the electron momentum and by the temperature of helium. 

Both ripplons and phonons are short-range scatterers, therefore the probability for an electron to scatter is independent of the state of other electrons, i.e., the electrons are scattered independently of each other. The scattering events are rare, whereas  the dynamics between the scattering events is fully controlled by the electron-electron interaction. This approach was used \cite{Moskovtsev2019} to describe self-diffusion and Wigner crystallization in the absence of a driving electric field. 

It should be emphasized that the self-diffusion in an interacting system does not give the long-wavelength diffusion coefficient. Therefore it does not give the electron mobility. This is particularly clear from noting that self-diffusion arises in an isolated electron system at a finite temperature, where the mobility is limited by the momentum transfer from the electrons to external scatterers, in our case, to ripplons and phonons in helium.

Direct simulation of scattering is essential to separate the momentum and energy relaxation in the system. For electrons on helium, the quasi-elastic ripplon scattering rate is at least an order of magnitude higher than that of inelastic scattering processes. Separating scattering mechanisms allows us to capture such effects as electron heating in the driven system. However, qualitatively, several effects of a periodic potential described below can also be reproduced in the standard molecular dynamics approach based on solving Langevin equations of motion with a friction force proportional to the electron velocity. The phenomenological friction coefficient has to be assumed small, so that the electron motion is strongly underdamped. 

To obtain reliable results we used a comparatively large system of 1600 electrons placed into a rectangular area, with periodic boundary conditions. The ratio of the sides of the rectangular area along the $y$ and $x$ axes was $L_y/L_x = 2/\sqrt{3}$, which allows a Wigner crystal with hexagonal symmetry to fit into the area \cite{Gann1979} (however, we studied the parameter range where Wigner crystallization did not occur). Importantly, we allowed the system to form a stationary distribution for a long time of $>10^6$ integration steps per electron and collected the data for more than $10^7$ steps. More details of the simulations are given in Appendix~\ref{sec:simulations}.   

\section{Mobility of a uniform system}
\label{subsec:uniform}

The conductivity of a strongly correlated electron system on helium is not described by the standard Boltzmann kinetic equation. In this equation, the momentum of an individual electron is randomized by successive short collisions with the vibrational excitations in helium and, for elastic scattering, the conductivity is given by the Drude expression $\sigma_{xx}=e^2n_s\langle\tau\rangle/m_e$, where $\langle\tau\rangle$ is the momentum relaxation time averaged over the Boltzmann distribution of the single-electron energy. 

We are not aware of a closed-form equation for the single-particle distribution function in the case of strong electron-electron interaction. In the electron liquid, after a collision with a vibrational excitation, and prior to the next collision, the momentum and energy of an individual electron are randomized by the electron-electron interaction. The long-wavelength many-electron conductivity can be calculated starting with the Kubo formula and using a transport equation for the many-electron Green function. In the case of elastic scattering this equation was derived earlier \cite{Dykman1979a,Dykman1997}, with the account taken of a strong effect of the electron-electron interaction on the scattering by vibrational excitations in the presence of a magnetic field. 

The analysis \cite{Dykman1997} immediately extends to the case of inelastic scattering.  The corresponding theory is developed in Appendix~\ref{sec:inelastic}. 

Even though the many-electron theory of magnetoconductivity was in an excellent agreement with the experiment with no adjustable parameters \cite{Lea1998}, the underlying picture of the many-electron relaxation could not be tested directly. In contrast, realistic simulations of the electron dynamics provide a means for such testing, as they allow one not only to calculate the (generally nonlinear) conductivity of the many-electron system, but also to find the electron energy distribution.
\begin{figure}[h]
	\includegraphics[scale=0.5]{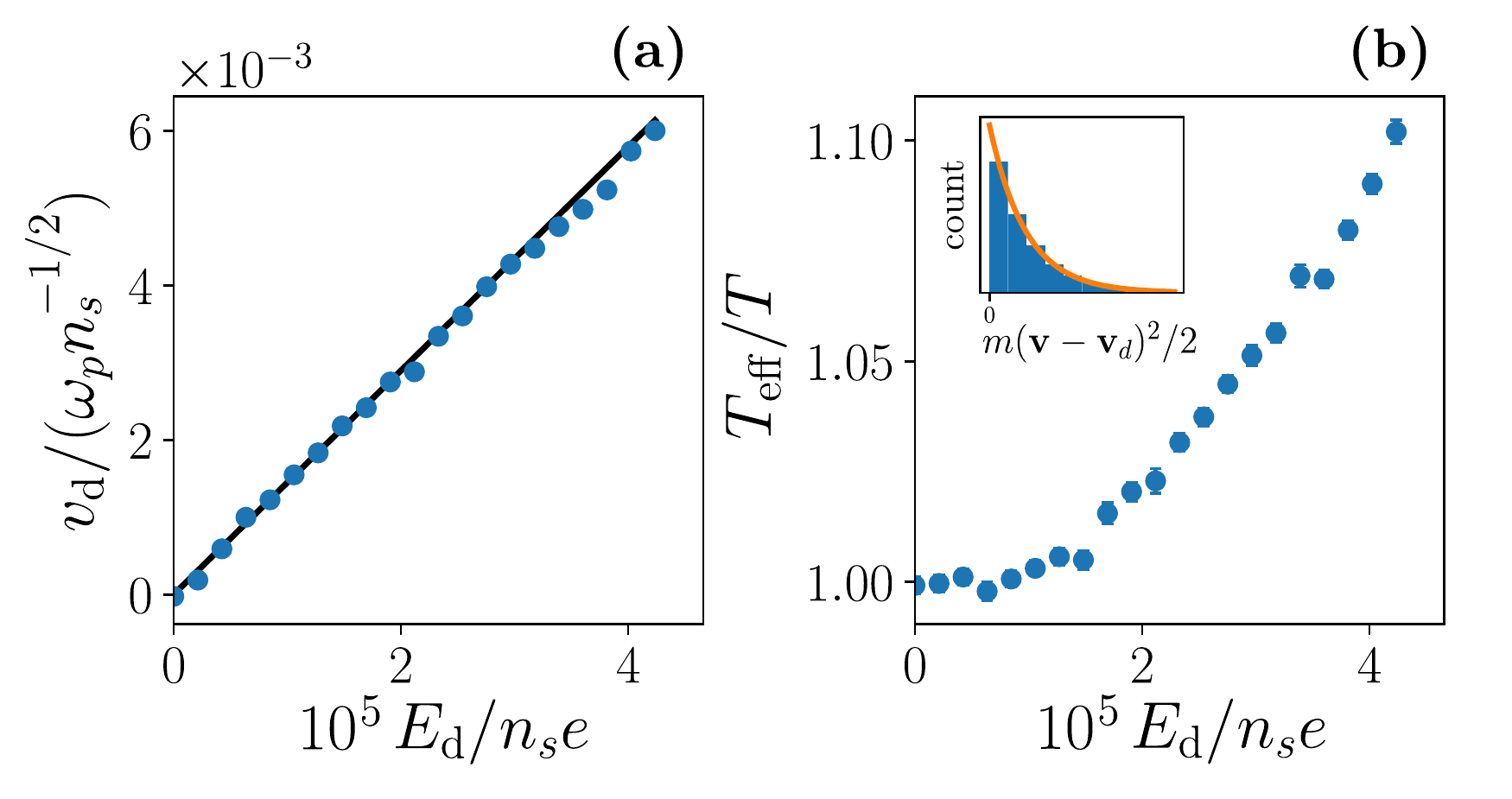}  
	\caption{The drift velocity $v_d$ (a) and the electron temperature  $T_{\rm eff}$ (b) vs the driving field $E_d$  for a uniform strongly correlated electron liquid. For $E_{\rm d}=0$, where the electron system is in thermal equilibrium with the excitations in helium, $\Gamma=90$. The filled circles are the simulation data. In (a) the black solid line is the velocity expected from the analytical expressions (D18) and (D19). The inset in (b) shows the distribution of the electron kinetic energies in the driven system measured in the frame moving with the drift velocity $\vb_d$ for $E_d/n_s e= 4.24\times 10^{-5}$; the solid line shows the Boltzmann distribution.}
	\label{fig:vd_uniform}
\end{figure}

The results of the simulations of the drift velocity of the spatially uniform electron liquid are presented in Fig.~\ref{fig:vd_uniform}~(a) and are compared with the theory.  The driving field $E_{\rm d}$ is scaled by the field $n_s e$, which is the field created by an electron at the distance on the order of the mean inter-electron distance; the studied fields $E_{\rm d}$  are much smaller than $n_s e$. The velocity is scaled by the factor $\omega_p n_s^{-1/2}$, which is the  velocity required to go over the characteristic interelectron distance over the time $1/\omega_p$, and again, $v_d\ll \omega_p n_s^{-1/2}$. 

For the parameters used in the simulations, the mobility $\mu=v_{\rm d}/E_{\rm d}$ for small $E_{\rm d}$ is $3.5\times 10^7$~cm$^2$/Vs. We note that the rate of quasielastic ripplon scattering is proportional to temperature. Therefore the value of $\mu$ depends not only on $\Gamma$, but also independently on $T$, with $\mu\propto 1/T$ for a fixed $\Gamma$. The above value and the plot in Fig.~\ref{fig:vd_uniform} refer to $T=0.354$~K, or equivalently, to $n_s \approx 1.15\times 10^8\,{\text cm}^{-2}$ for $\Gamma=90$. Strictly speaking, for such $n_s$ one should take into account the coupling to ripplons due to the field that presses the electrons against the helium surface \cite{Andrei1997}; this would reduce the numerical value of the mobility, but will not affect the qualitative results discussed below.

The simulations demonstrated that the distribution over the electron energy is the Boltzmann distribution with the effective temperature $T_{\rm eff}$ that increases with the driving field. This increase is shown in Fig.~\ref{fig:vd_uniform}~(b). Such form of the distribution confirms the underlying assumption of the analytical theory. For the fields $E_{\rm d}$ used in the simulations, the electron heating is weak and has a negligible effect on the mobility. 

\section{Electron liquid in a periodic potential}
\label{sec:mobility_commensurate}

The effect of the periodic potential on the electron system strongly depends on the interrelation between the period of the potential $2\pi/Q$ and the mean interelectron distance. More specifically, one can think of the electrons forming a Wigner crystal with a triangular lattice and  define the commensurability parameter
\begin{align}
\label{eq:p_parameter}
p_c = (\sqrt{3}/2n_s)^{1/2}Q/2\pi \equiv a_s\sqrt{3}Q/4\pi,
\end{align}
where $a_s =(2/\sqrt{3}n_s)^{1/2}$ is the interelectron distance in the Wigner crystal.
  
For the considered one-dimensional potential,  $p_c$ gives the ratio of the distance between the rows of the Wigner crystal $a_s\sqrt{3}/2$ and the period of the potential. The rows here have been chosen in such a way as to minimize the inter-electron distance in a row, see Appendix~\ref{sec:Gaussian_confinement}. Depending on $p_c$, the periodic potential can lead to Wigner crystallization for smaller values of the plasma parameter $\Gamma$ (\ref{eq:parameters}), i.e., for higher temperatures than in a uniform system, or can impede crystallization \cite{Moskovtsev2019}. 

\begin{figure}[h]
	\includegraphics[scale=0.5]{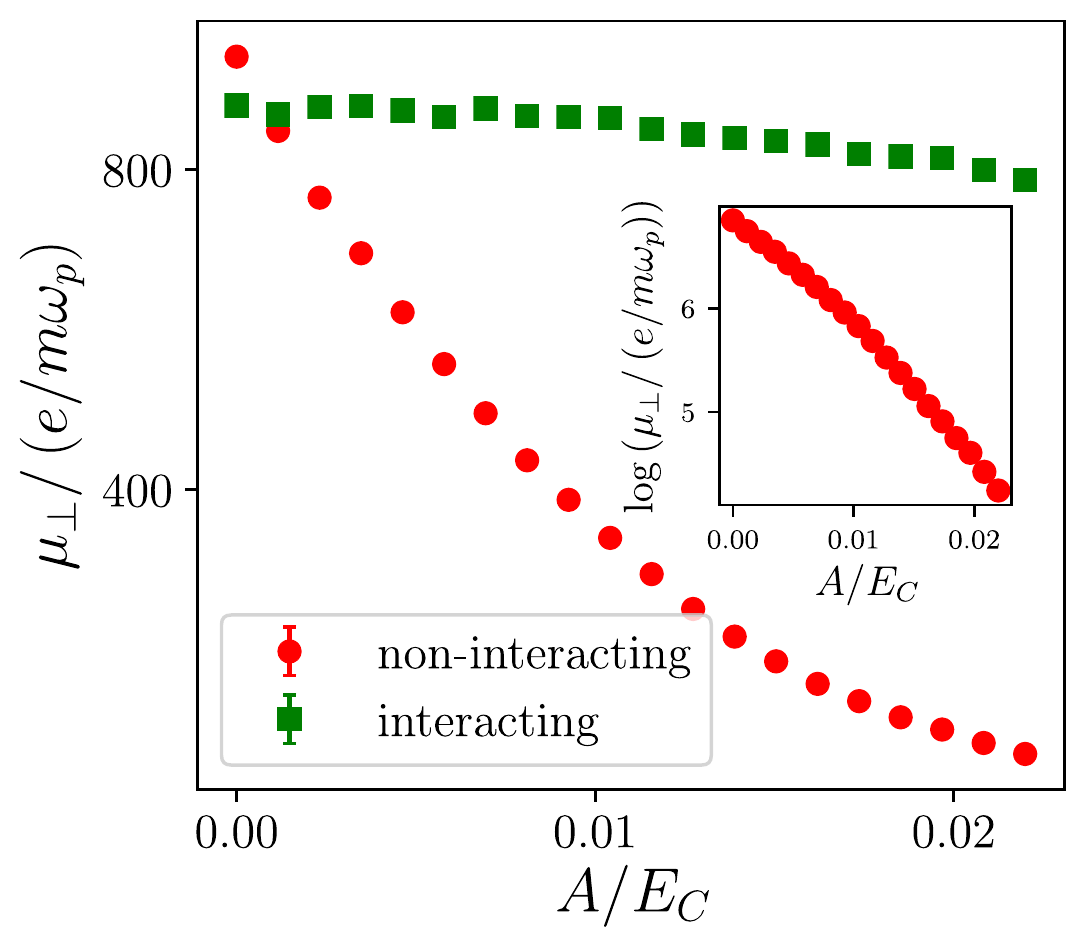} 
	\caption{The mobility $\mu_{\perp}\equiv \mu_{xx}$ transverse to the troughs of the sinusoidal periodic potential (\ref{eq:Hamiltonian_ee}) with amplitude $A$ scaled by $\mu_{\rm eff} = e/m\omega_p$, see Eq.~(\ref{eq:diff_unit}). The results refer to $\Gamma=90$ and the commensurability parameter $p_c=0.3$. The squares and the circles show the data for the electron liquid and the ideal electron gas, respectively.}
	\label{fig:mu_incommensurate}
\end{figure}

The effect of a weak potential  on crystallization is small if $p_c$ is such that a crystal has to be strongly distorted to fit into the periodic  potential \cite{Bak1982}. For a classical Wigner crystal, the effect is also small if $p_c$ is small, as  the electron-electron interaction effectively screens the potential. Respectively, one expects that not only the mobility along the potential troughs, but also the mobility transverse to the troughs $\mu_\perp$ will be weakly affected for small $p_c$. This is indeed seen in Fig.~\ref{fig:mu_incommensurate}. 

As mentioned in the Introduction, the mobility of the electron liquid in a periodic potential should strongly differ from that of an ideal electron gas. This is also seen in Fig.~\ref{fig:mu_incommensurate}. The scale of the barrier height $2A$ on which the single-electron transverse mobility $\mu_\perp$ exponentially falls off should be given by the temperature. In Fig.~\ref{fig:mu_incommensurate},  on the abscissa, the scaled amplitude of the periodic potential $A/E_C$ can be written as $\Gamma^{-1} (A/k_BT)$, and as expected, for the studied $\Gamma=90$ the exponential fall-off of $\mu_\perp$ with $A$ starts for $A/E_C\gtrsim 0.01$. 

\section{``Maximally commensurate'' potential}
\label{subsec:commensurate}

The effect of the periodic potential is most pronounced in the case where $p_c=1$. In this case, if the electrons formed a crystal, it would be fully commensurate with the potential, cf Appendix~\ref{sec:Gaussian_confinement}. We call such a potential maximally commensurate, with one electron row per one potential trough. Even where the amplitude of the potential $A$ is much smaller than the electron correlation energy $E_C$ and the temperature is well above the temperature of the Wigner crystallization in a uniform system,  the electron liquid preferentially occupies the potential troughs. Respectively, the electron density can be comparatively strongly periodically modulated.

The strong effect of the maximally commensurate potential comes from the fact that the potential lifts the translational and orientational  symmetries of the electron liquid and, rather than competing with the electron correlations, it  constructively interferes with them. Therefore such a potential can significantly modulate the electron density without changing the small-amplitude fluctuations about quasi-equilibrium electron positions in the liquid. 

The latter can be seen from the following argument. A single electron localized at the minimum of a potential trough vibrates normal to the trough with frequency $(AQ^2/m)^{1/2}$. On the other hand, an electron in an unconfined electron liquid vibrates about its quasiequilibrium position with frequency $\sim \omega_p$ (\ref{eq:parameters}).  For $p_c=1$, the ratio of the squares of these frequencies is $(A/E_C)4\pi^{3/2}/\sqrt{3}$. In the range of the potential strengths and electron densities that we studied this ratio was very small (it is equal to $\approx 0.02$ for $A/E_C = 0.0016$). 

By the same token, a maximally commensurate potential may be expected to weakly affect the relative positions of the electrons. Those are characterized by the two-particle correlation function. For a system in a periodic potential it can be defined as
\begin{align}
\label{eq:pair_correlator_full}
g^{(2)}_U(\rb',\rb'') =\frac{1}{\rho_s(\rb')\rho_s(\rb'')}
\sumprime{n,m}\delta(\rb'-\rb_n)\delta(\rb''-\rb_m),
\end{align}
where  $\rho_s(\rb)$ is the electron density. For $\rho_s(\rb)=$~const, equation~(\ref{eq:pair_correlator_full}) goes over into the standard expression for the pair correlation function of a spatially uniform system. 

For the considered potential $U(\rb)$, the density $\rho_s(\rb)$ is periodic in $x$ with period $2\pi/Q$ and is independent of $y$; its average value is $n_s$,
\begin{align}
\label{eq:density_periodic}
\rho_s(\rb) = n_s\left[1+\sum_m\alpha_m\cos(mQx)\right].
\end{align}
The coefficients $\alpha_m$ quickly fall off with the increasing $m$ for a weak potential, where they can be found by a perturbation theory in $A/k_BT$. They can be found also if the electrons are strongly confined within the troughs, see Appendix~\ref{sec:Gaussian_confinement}.

\begin{figure}[t]
	\includegraphics[scale=0.45]{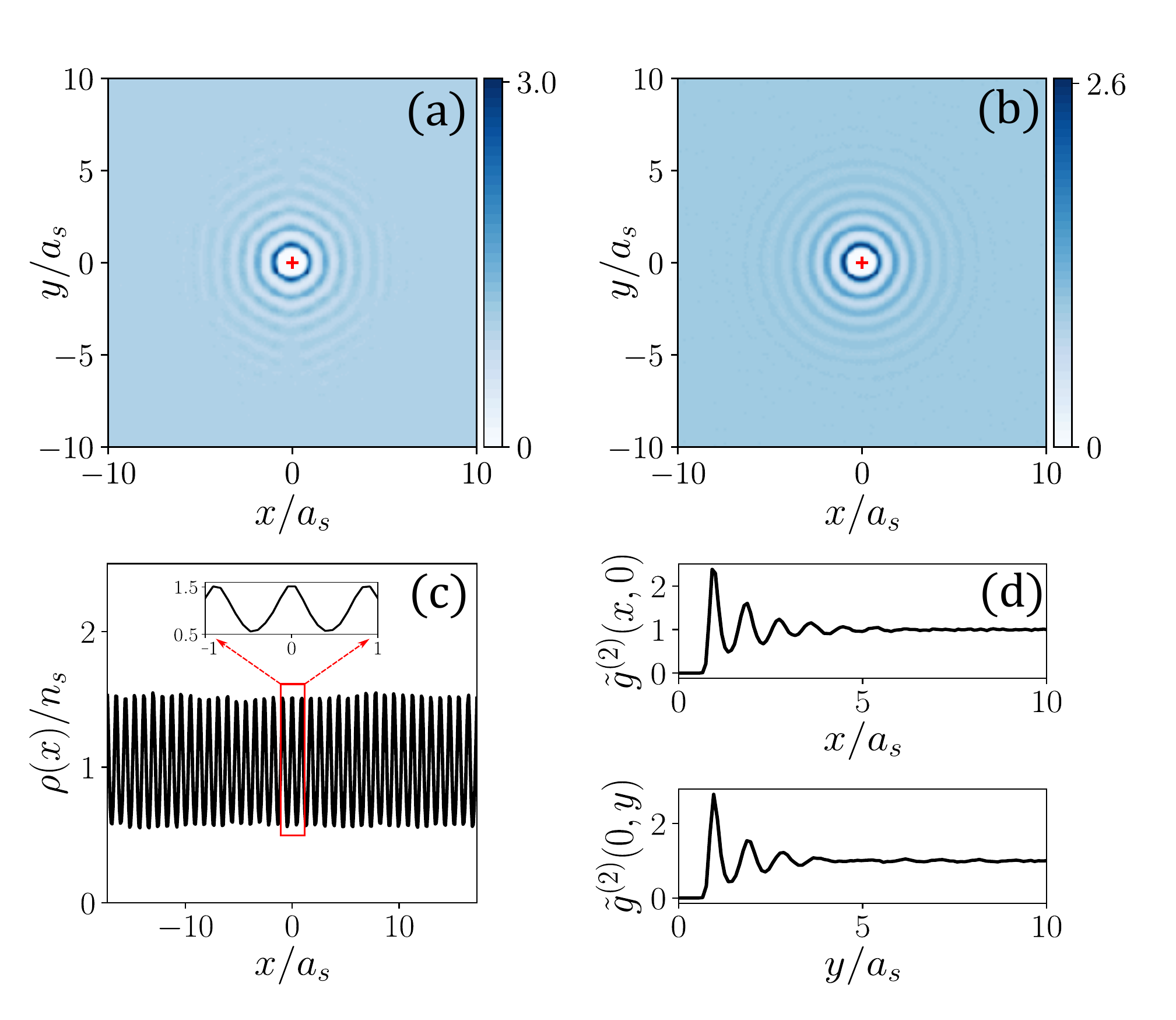}
	\caption{(a) The pair correlation function $\tilde{g}^{(2)}(\rb)$, Eq.~(\ref{eq:pair_correlation}), for the maximally  commensurate periodic potential, $p_c=1$. The scaled amplitude of the potential is $A/E_C=0.00157$, and the plasma parameter is $\Gamma=90$. (b) The pair correlation function in the absence of a potential for the same $\Gamma=90$. (c) The averaged over time electron density along the direction $x$ of the oscillations of the potential for the same parameters as in (a). (d) One-dimensional slices of the pair correlation function  $\tilde{g}^{(2)}(\rb)$ in panel (a).}
	\label{fig:pair_distribution}
\end{figure}

An example of $\rho_s(\rb)$ is shown in Fig.~\ref{fig:pair_distribution}. The result refers to a small ratio of the potential amplitude to temperature, $A/k_BT =\Gamma A/E_C \approx 0.14$. The modulation of $\rho_s(\rb)$ is much stronger than it would be  in the single-electron picture ($\propto \exp[-(A/k_BT) \cos Qx]$). This is the result of the commensurability: the potential constructively interferes with the strong  electron correlations and just lifts the translational and rotational degeneracy of the system.

The function $g^{(2)}_U(\rb',\rb'')$ defined by Eq.~(\ref{eq:pair_correlator_full}) depends on $y'-y''$ and does not change if both $x'$ and $x''$ are incremented by the period of the potential $2\pi/Q$. Therefore, if written as a function of $x'-x''$ and $(x'+x'')/2$, it is a periodic function of $(x'+x'')/2$. In addition, $g^{(2)}_U(\rb',\rb'')\to 1$ for $|x'-x''|,|y'-y''|\to \infty$. 

One can also consider the zeroth Fourier component of   $g^{(2)}_U(\rb',\rb'')$ with respect to $(x'+x'')/2$ and introduce the function
\begin{align}
\label{eq:pair_correlation}
&\tilde{g}^{(2)}(\rb) = \frac{n_s}{N}\int d\Rb g^{(2)}_U\left(\Rb+\frac{1}{2}\rb,\,\Rb - \frac{1}{2}\rb\right)\nonumber\\ 
&=
\frac{n_s}{N}\sumprime{n,m}\delta[\rb - (\rb_n-\rb_m)]/[\rho_s(\rb_n)\rho_s(\rb_m)],
\end{align}
where $N$ is the total number of electrons.
Function $\tilde{g}^{(2)}(\rb)$ is an analog of the pair correlation function of a uniform system. 

The correlation function  $\tilde{g}^{(2)}(\rb)$ is shown in Fig.~\ref{fig:pair_distribution}. Remarkably, even though the electron density is strongly periodically modulated even by the weak potential used in the calculation, $\tilde g^{(2)}$ is very similar to the pair correlation function of a spatially uniform system for the same $\Gamma$, which is also shown in Fig.~\ref{fig:pair_distribution}. Both with and without the potential, the correlations in the electron liquid decay over several interelectron distances.

\begin{figure}[h]
	\includegraphics[scale=0.27]{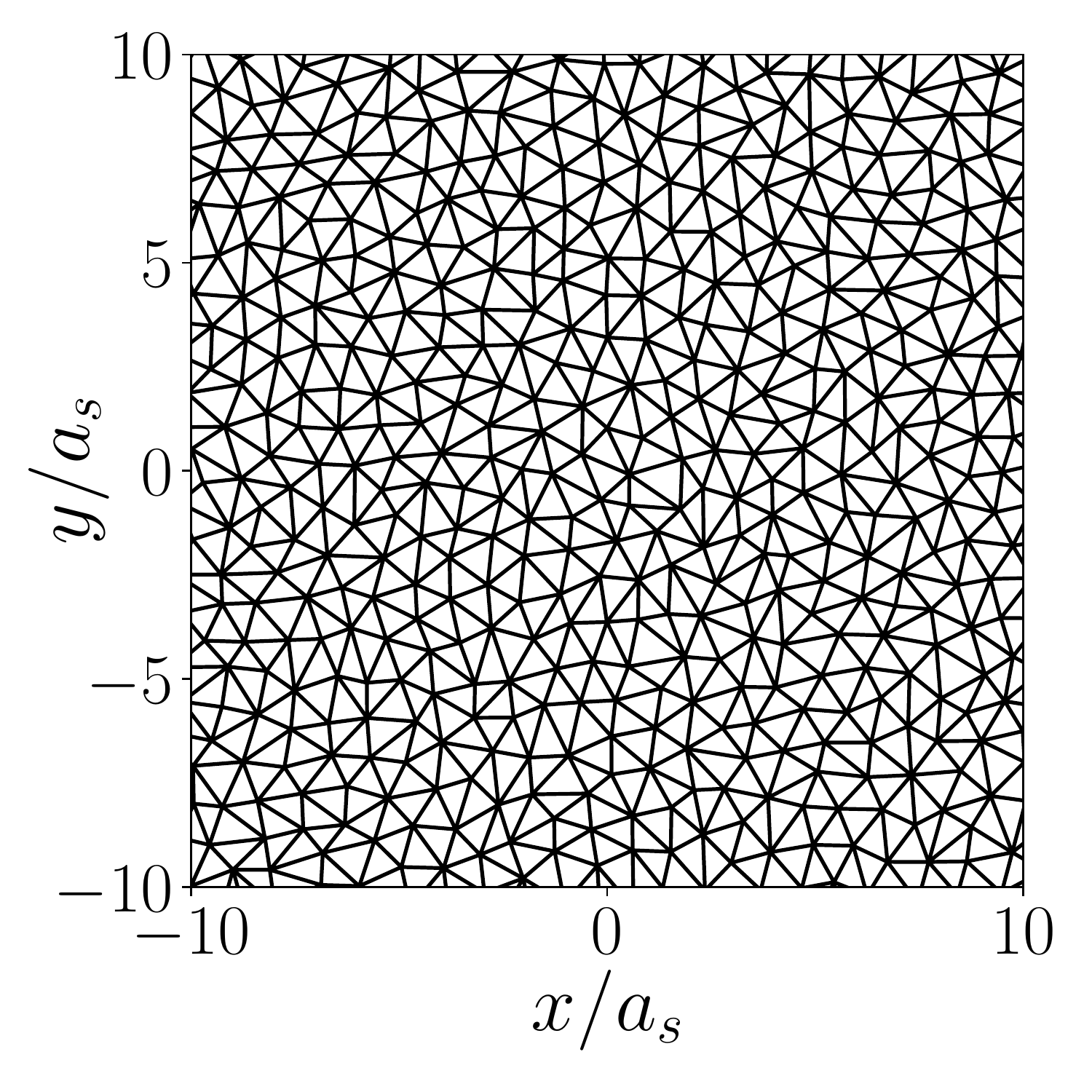}~
	\includegraphics[scale=0.27]{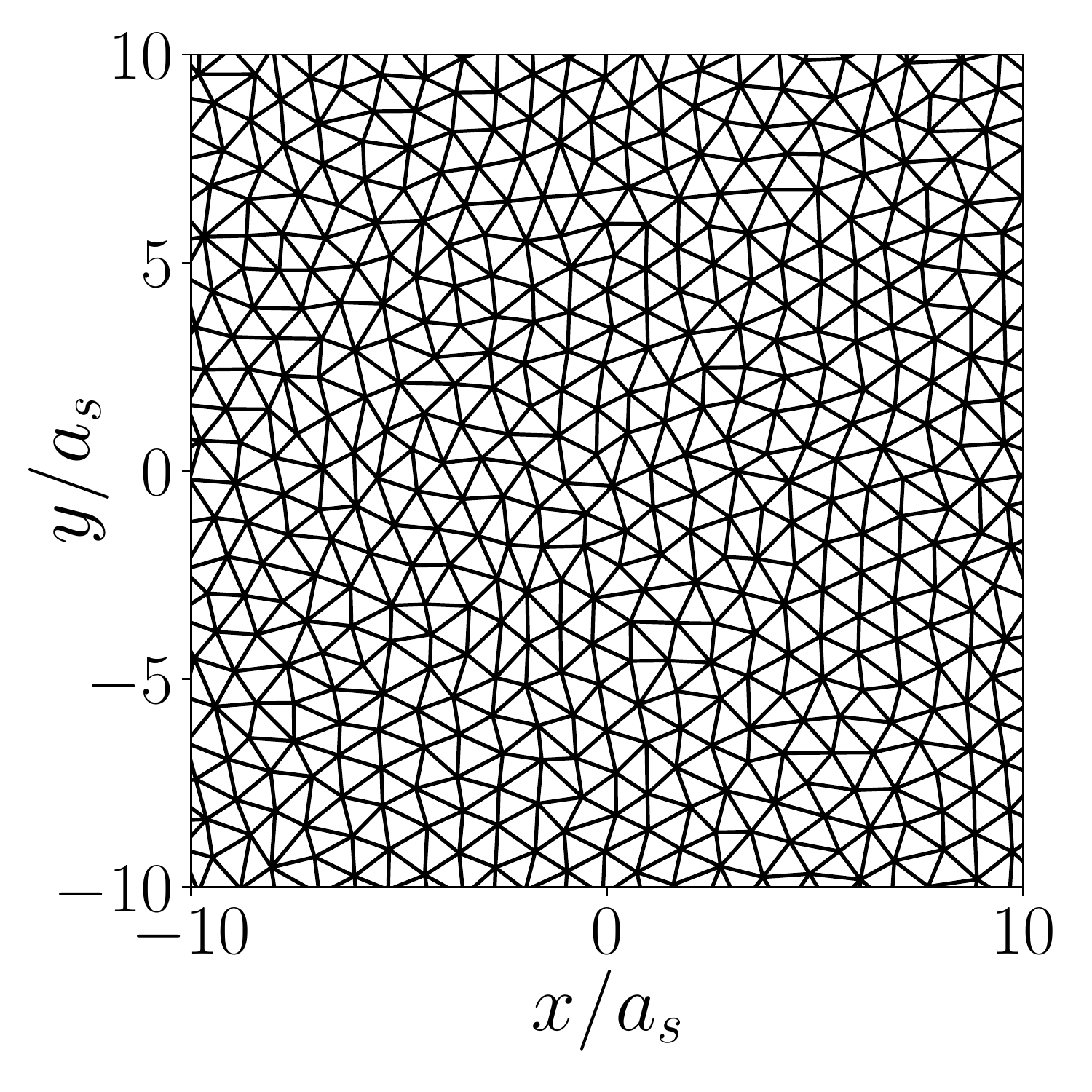}  
	\caption{Delaunay triangulation for a snapshot of the electron liquid in the periodic potential of amplitude $A/E_C=0.00157$; $p_c=1$. The electrons are represented by vertices. The plasma parameter is $\Gamma = 50$ and $90$ for the left and right panels, respectively.}
	\label{fig:delaunay}
\end{figure}

Decay of electron correlations can be also seen from the Delaunay triangulation shown in Fig.~\ref{fig:delaunay}. The short-range order is clearly seen for the both values of the plasma parameter $\Gamma$. For  larger $\Gamma$, the anisotropy of the density associated with the alignment along the potential troughs (along the $y$-axis) is visible, and also the number of unbound vortices with 5 or 7 nearest neighbors is smaller. As $\Gamma$ increases, such defects play an increasingly important role in the electron dynamics. We note that the periodic density modulation shown in Fig.~\ref{fig:pair_distribution} is not at all obvious from Fig.~\ref{fig:delaunay}, although the very  compressibility of the electron liquid is clearly seen. As mentioned above, the results of Fig.~\ref{fig:pair_distribution} were obtained by averaging over many frames of the type of those in Fig.~\ref{fig:delaunay}.

\subsection{Electron mobility transverse to the potential troughs}

Placing the electron system into a sufficiently strong commensurate periodic potential should strongly affect the electron mobility. In particular, the mobility transverse to the potential troughs $\mu_\perp$ should decrease. This decrease should sensitively depend on the amplitude of the potential and the temperature. In turn, as we show, this dependence may be used to reveal and characterize the correlations in the system. 

It is instructive to compare the mobility $\mu_\perp$ with the self-diffusion coefficient. For a spatially uniform classical 2D electron liquid the self-diffusion coefficient is of the order of $D_0=k_BT/m\omega_p$ \cite{Moskovtsev2019}, an estimate close to the De Gennes estimate for a normal three-dimensional liquid \cite{DeGennes1959}. In our simulations we used a fixed electron density and varied the temperature. Then one can scale the self-diffusion coefficient and the mobility by the temperature-independent parameters $D_{\rm eff}$ and $\mu_{\rm eff}$, respectively, 
\begin{align}
\label{eq:diff_unit}
D_{\rm eff} = \omega_p/n_s = 2\sqrt{\pi}\,\Gamma D_0, \qquad \mu_{\rm eff} = e/m\omega_p.
\end{align}
For the electron liquid in the absence of a periodic potential and far from the crystallization transition, we expect $D\sim D_0 \ll D_{\rm eff}$ and  $\mu_\perp \sim e\tau/m\gg \mu_{\rm eff}$, where $\tau$ is the electron relaxation time due to the scattering by ripplons, $\tau\gg \omega_p^{-1}$.

\begin{figure}[t]
	\includegraphics[scale=0.4]{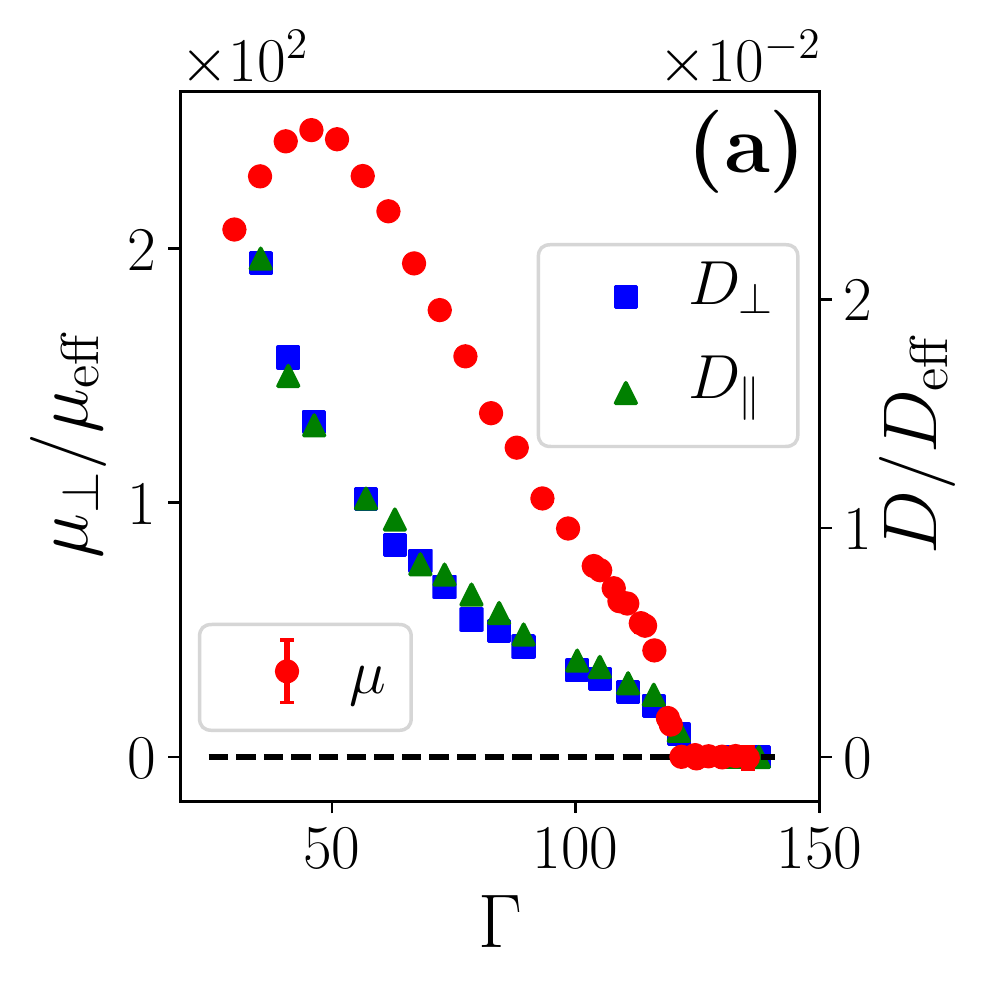}\hfill
\includegraphics[scale=0.4]{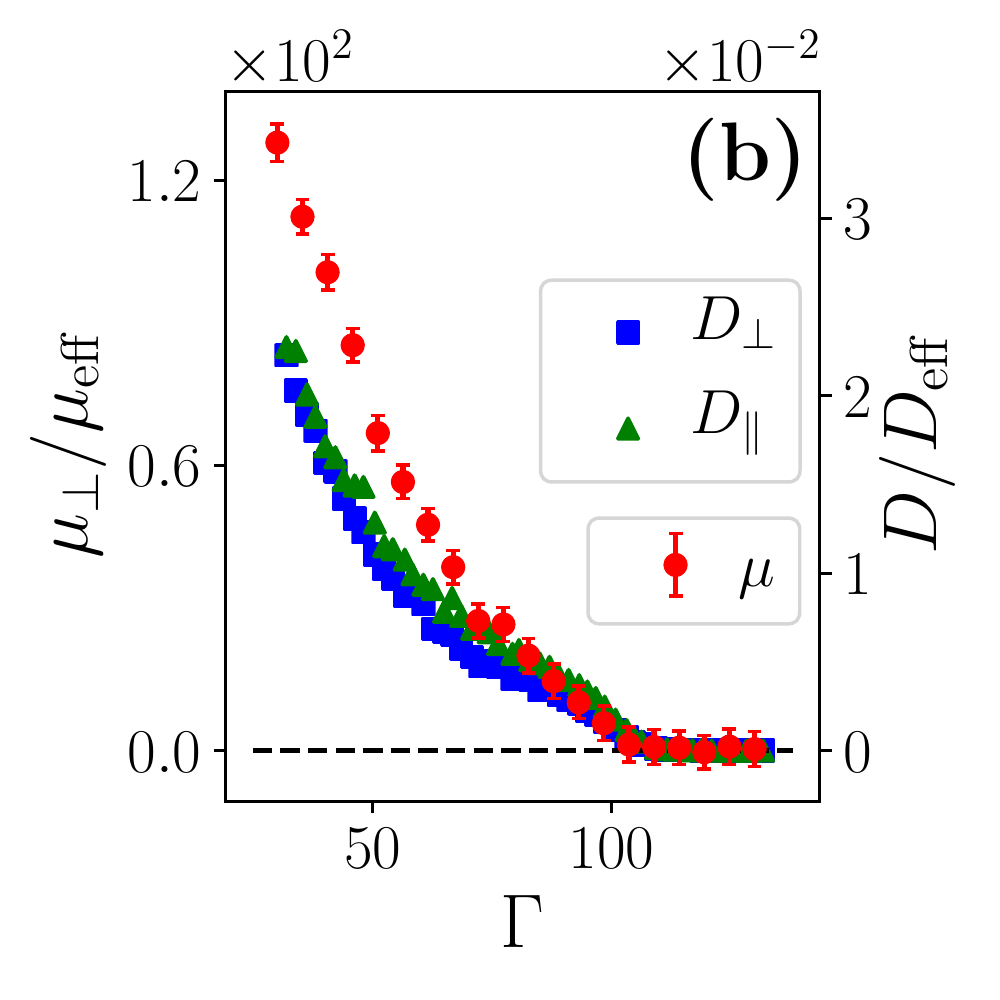}
	\caption{The mobility of the electron liquid $\mu_{\perp}$ transverse to the troughs of the periodic potential (left $y$-axis) and the coefficients of self-diffusion along ($D_\parallel$) and transverse ($D_\perp$) to the troughs (right $y$-axis) as functions of the reciprocal temperature $T^{-1}\propto \Gamma$ (the electron density is fixed). The results refer to the maximally commensurate potential, $p_c=1$. The amplitudes of the potential in the left and right panels are $A/E_C=0.000628$ and $0.00157$, respectively.}
	\label{fig:mu_D_vs_T}
\end{figure}

When the electron liquid is in a 1D periodic potential, both the mobility and the self-diffusion become anisotropic. Similar to the case of the mobility, one can introduce the coefficients of self-diffusion along and transverse to the troughs, $D_\parallel$ and $D_\perp$, respectively. They are defined by the long-time mean-square displacements of an electron in the corresponding directions, $ \langle [y_n(t)-y_n(0)]^2\rangle = 2D_\parallel t$ and $ \langle [x_n(t)-x_n(0)]^2\rangle = 2D_\perp t$. Simulations of these coefficients are described in Ref.~\onlinecite{Moskovtsev2019}.

The mobility along the troughs is limited by the scattering by ripplons and should be weakly affected by the potential. We have indeed seen this in the simulations and do not discuss this mobility. In contrast, self-diffusion is controlled by the electron correlations, and therefore the coefficient of self-diffusion along the troughs $D_\parallel$ is strongly affected by the commensurate potential. Moreover, when the system freezes into a Wigner crystal, the self-diffusion along the troughs vanishes \cite{Moskovtsev2019}, whereas the mobility along the troughs  does not.  

\begin{figure}[h]
	\centering
	\includegraphics[scale=0.48]{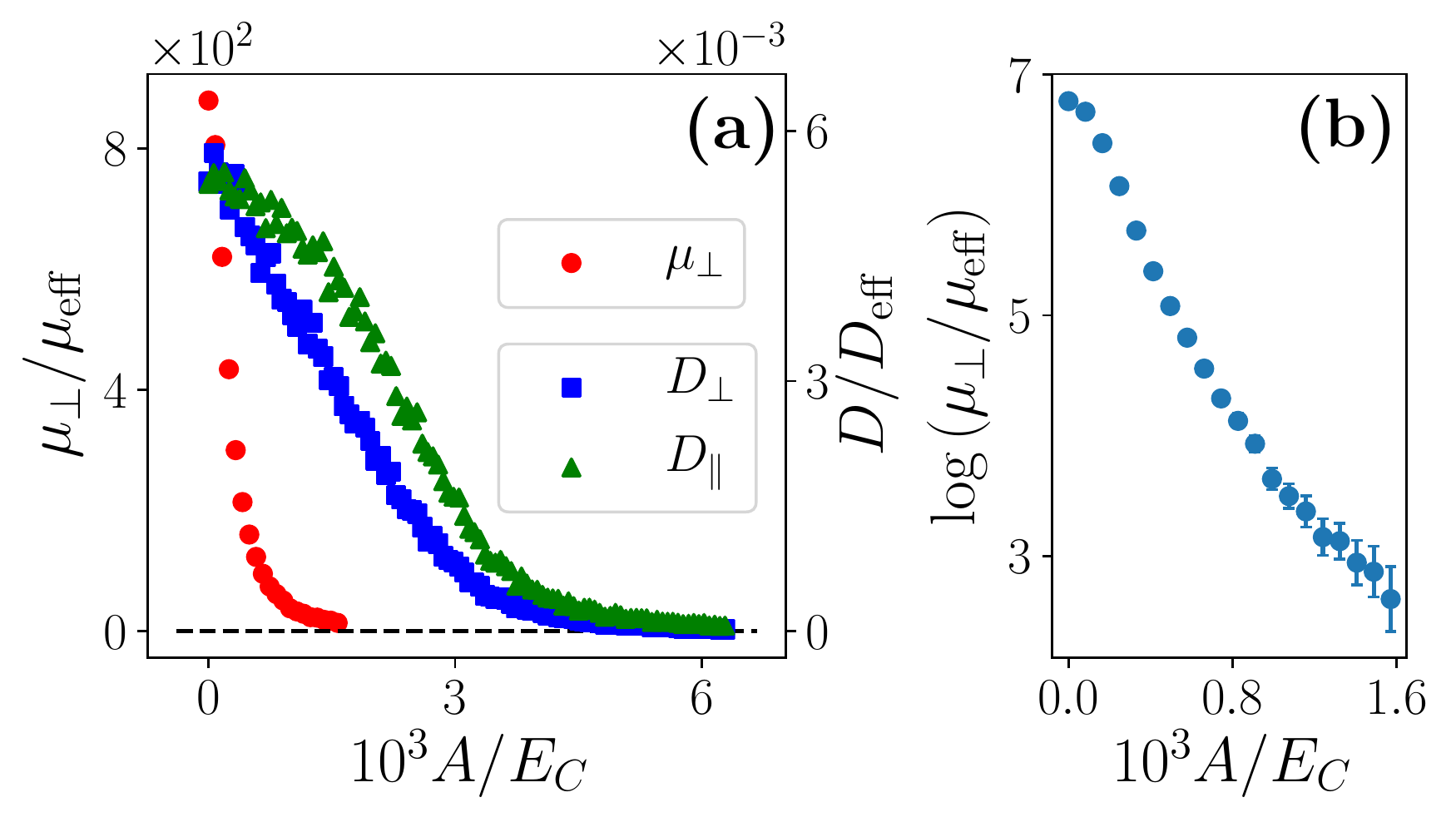}
	\caption{(a) The mobility of the electron liquid $\mu_{\perp}$ transverse to the potential troughs (left $y$-axis) and the coefficients of self-diffusion along ($D_\parallel$) and transverse ($D_\perp$) to the troughs (right $y$-axis) as functions of the potential amplitude. The results refer to the maximally commensurate potential, $p_c=1$, and to the plasma parameter $\Gamma = 90$. Note the difference in the scales compared to Fig.~\ref{fig:mu_D_vs_T}(b). (b) The logarithm of the transverse mobility for the same $\Gamma=90$.}
	\label{fig:mu_D_vs_A}
\end{figure}

In Fig.~\ref{fig:mu_D_vs_T},  we show the dependence of the mobility transverse to the troughs $\mu_\perp$  and the self-diffusion coefficients $D_\parallel$ and $D_\perp$ on temperature for two values of the potential amplitude $A$. In Fig.~\ref{fig:mu_D_vs_A}, we show the dependence of these parameters on $A$ for a fixed temperature.  Self-diffusion monotonically decreases with the decreasing temperature (increasing $\Gamma$) and with the increasing potential amplitude.  Moreover, the  self-diffusion coefficients  $D_\parallel$ and $D_\perp$  are close to each other  for the considered weak potential in Fig.~\ref{fig:mu_D_vs_T}, $A\ll E_C$ (they become very different for a stronger potential \cite{Moskovtsev2019}). 

In contrast, for a very weak potential, Fig.~\ref{fig:mu_D_vs_T}(a), the mobility depends on $T$ nonmonotonically, first increasing  with the decreasing $T$ for higher temperature. This increase is related to the decrease of the rate of  electron scattering by ripplons, which is  $\propto T$. However, for still smaller $T$, the mobility decreases with the decreasing $T$ (increasing $\Gamma$).

Figure \ref{fig:mu_D_vs_A} shows that $\mu_\perp$ monotonically decreases with the increasing potential amplitude $A$. This decrease is close to exponential. Moreover, for the value of the plasma parameter $\Gamma=90$ shown in the figure, the mobility falls off  with the increasing $A$ sharper than the self-diffusion coefficients. 

Both the transverse mobility and the self-diffusion should vanish where electrons crystallize in a commensurate potential.  The mobility is advantageous for finding the Wigner crystallization temperature as it is much easier to access in the experiment then the self-diffusion. However, measuring a very small mobility is complicated not only in the experiment, but also in the simulations, as seen from the error bars in Fig.~\ref{fig:mu_D_vs_A}.

We note the difference in the scales in Figs.~\ref{fig:mu_D_vs_T}(b) and \ref{fig:mu_D_vs_A}(a). The scale for $\mu_\perp$ is larger whereas the scale for $D_\parallel$ and $D_\perp$ is smaller in  \ref{fig:mu_D_vs_A}(a). On the scale of Fig.~\ref{fig:mu_D_vs_T}, the difference between the parameter values where the mobility and the self-diffusion become close to zero is within the error bars. 

To gain a better feeling for the scaling factors in Figs.~\ref{fig:mu_D_vs_T} and \ref{fig:mu_D_vs_A}, it is instructive to look at the Einstein ratio $\mu_\perp k_BT/eD$, which in the single-electron approximation should be equal to one. For our scaling factors we have 
\begin{align}
\label{eq:Einstein}
\mu_{\rm eff} k_BT (eD_{\rm eff})^{-1}=(2\sqrt{\pi}\,\Gamma)^{-1},
\end{align}
i.e., for $\Gamma= 10^2$ we have $\mu_{\rm eff} k_BT/eD_{\rm eff}\approx 1/300$. Therefore for $\mu_\perp$ and $D$ measured in the units of $\mu_{\rm eff}$ and $D_{\rm eff}$, if the Einstein relation held, the Einstein ratio would be $\sim 300$. Instead, as seen from  Figs.~\ref{fig:mu_D_vs_T} and \ref{fig:mu_D_vs_A}, it is $\sim 10^4$. This shows that, even though the mobility is reduced by the periodic potential, it is still orders of magnitude higher than what follows from the Einstein relation. We emphasize again that $D_\perp$ and $D_\parallel$ are the coefficients of self-diffusion, not of the long-wavelength diffusion, therefore the Einstein relation should not hold; it is just that the difference with this relation is as large as two orders of magnitude.

\subsection{Many-electron nature of overcoming the barriers}

\begin{figure}[h]
	\centering
	\includegraphics[width=0.3\textwidth]{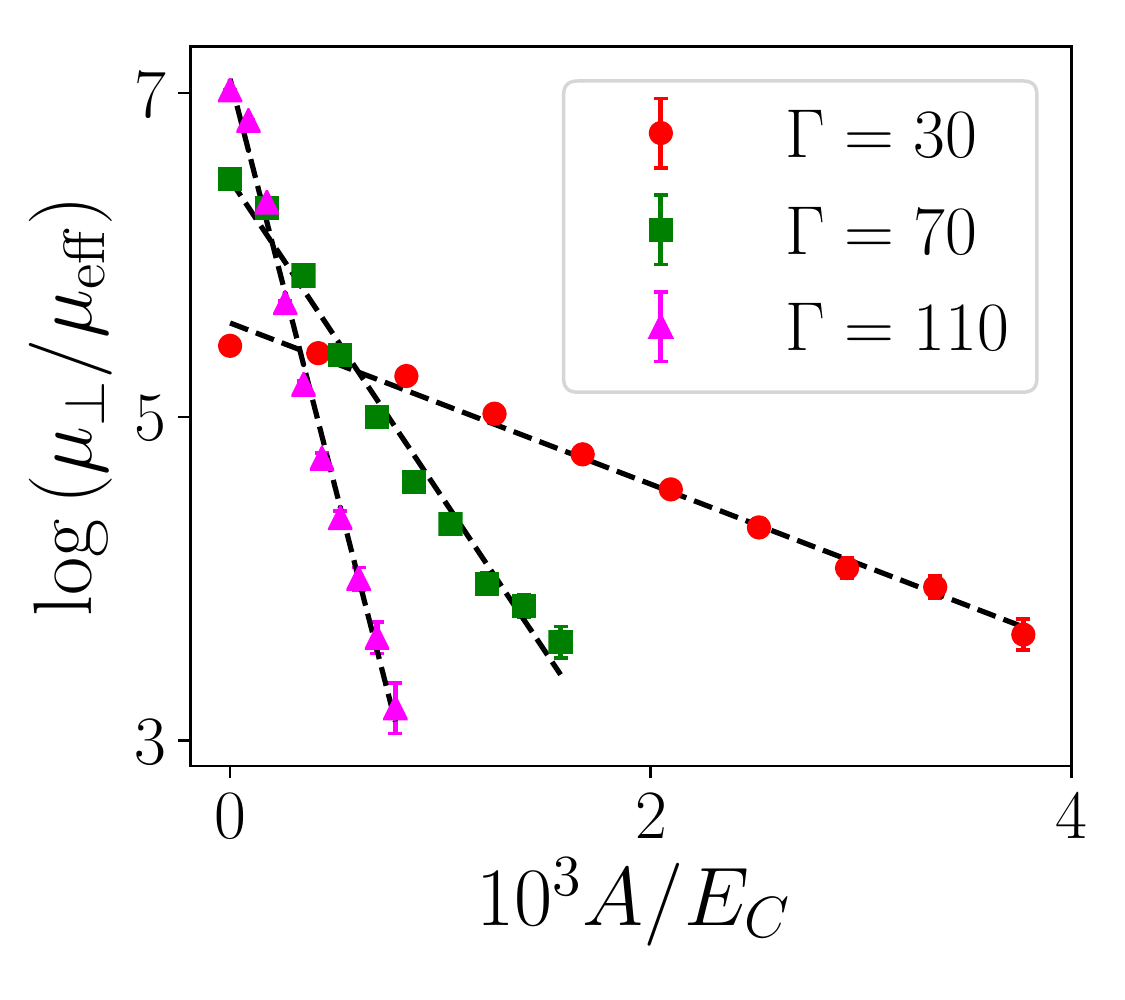}  
	\caption{The mobility $\mu_{\perp}$ of the strongly correlated electron system as a function of the potential amplitude $A$ for three values of $\Gamma$. The dashed lines show the least squares fit. }
	\label{fig:mu_vs_A_three}
\end{figure}

The transverse drift in the commensurate potential involves overcoming the periodically repeated potential barriers. One can picture the transport of the electron liquid in a small electric field along the $x$-axis as motion in a slightly tilted washboard potential. If the electron system were incompressible and formed a Wigner crystal, it would not move in the commensurate potential, as such motion would involve overcoming a barrier by all electrons at a time, i.e., the barrier height would be proportional to the system size. The very occurrence of the mobility is thus due to the correlation length of the electron system being finite. Still one would expect that the mobility will exponentially depend on the barrier height. The corresponding dependence has been indeed seen in the simulations and is shown in Fig.~\ref{fig:mu_vs_A_three} for several values of $\Gamma$.

To understand the result one can think of a cartoon of the electron liquid as made up of ``clusters'' with a typical size given by the correlation length $\xi$. The electrons are reasonably well ordered within clusters. Such clusters can be seen in Fig.~\ref{fig:delaunay}, particularly distinctly for $\Gamma=90$. The number of electrons in a cluster is $\sim \pi (\xi/a_s)^2$. The mobility results from the electron motion in which the clusters rearrange and move while the correlation length is preserved. For the considered case of maximal commensurability, the ultimate result is that the areas with a typical size given by the correlation length go over the potential barriers of height $2A$.  Such transitions are thermally activated. Thus the mobility should contain the Arrhenius factor,
\begin{align}
\label{eq:correlated_mobility}
\mu_\perp \propto \exp(-\gamma A/E_C), \qquad \gamma\sim  2\pi\,\Gamma \xi^2/a_s^2.
\end{align} 
We note that $\Gamma A/E_C \equiv A/k_BT$, so that the exponent in $\mu_\perp$ is $\gamma A/E_C \sim (2\pi \xi^2/a_s^2)A/k_BT$. It is important that, although the mobility is activated, $\gamma \propto \Gamma\propto 1/T$, the overall dependence of $\gamma$ on temperature is more complicated, because the correlation length depends on temperature. This leads to a nonexponential dependence of $\mu_\perp$ on $1/T$. Such a dependence has been seen in our simulations and can be inferred from Fig.~\ref{fig:mu_D_vs_T}.  

\begin{figure}[h]
	\centering
	\includegraphics[width=0.35\textwidth]{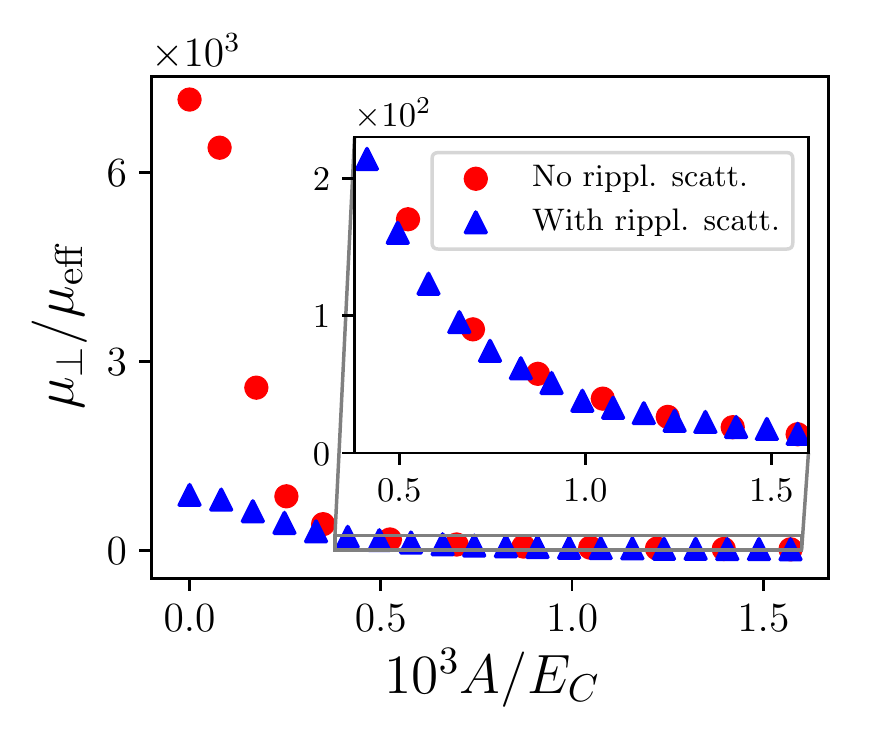} 
	\caption{A comparison of the mobility $\mu_{\perp}$ of the electron liquid as a function of the potential amplitude $A$ with and without electron scattering by ripplons.  The results refer to $\Gamma=90$. The inset shows the results in the region $A/E_C > 3.8\times 10^{-4}$, where the mobility is described by the activation law (\ref{eq:correlated_mobility}). Notice the different scales of $\mu_\perp$ in the main figure and the inset.}
	\label{fig:no_ripplons}
\end{figure}

The picture of a locally nonuniform transport implies internal friction in the electron liquid, with the momentum of the system as a whole being ultimately transferred to the periodic potential. If this is indeed the case, the mobility $\mu_\perp$ in the regime where it is activated, Eq.~(\ref{eq:correlated_mobility}), should become largely independent of the ripplon scattering, which controls the mobility in the absence of the potential. This is indeed seen from Fig.~\ref{fig:no_ripplons}.

It is important that, for the potential amplitudes we are studying, the correlation energy is much larger than the amplitude of the potential. Therefore the correlation length should be approximately the same as in the absence of the potential, the  picture corroborated by Fig.~\ref{fig:pair_distribution}. We determined the correlation length approximately from the radial distribution function  $g(r)$ of the electron liquid in the absence of the potential,
\begin{align}
\label{eq:radial_distribution}
g(r) = (2\pi r n_s N)^{-1}\sumprime{n,m}\delta(r-|\rb_n-\rb_m|).
\end{align}
This expression immediately follows from Eq.~(\ref{eq:pair_correlation}) if $\rho_s(\rb) = n_s$ and the system is isotropic.

\begin{figure}[t]
\includegraphics[width=0.23\textwidth]{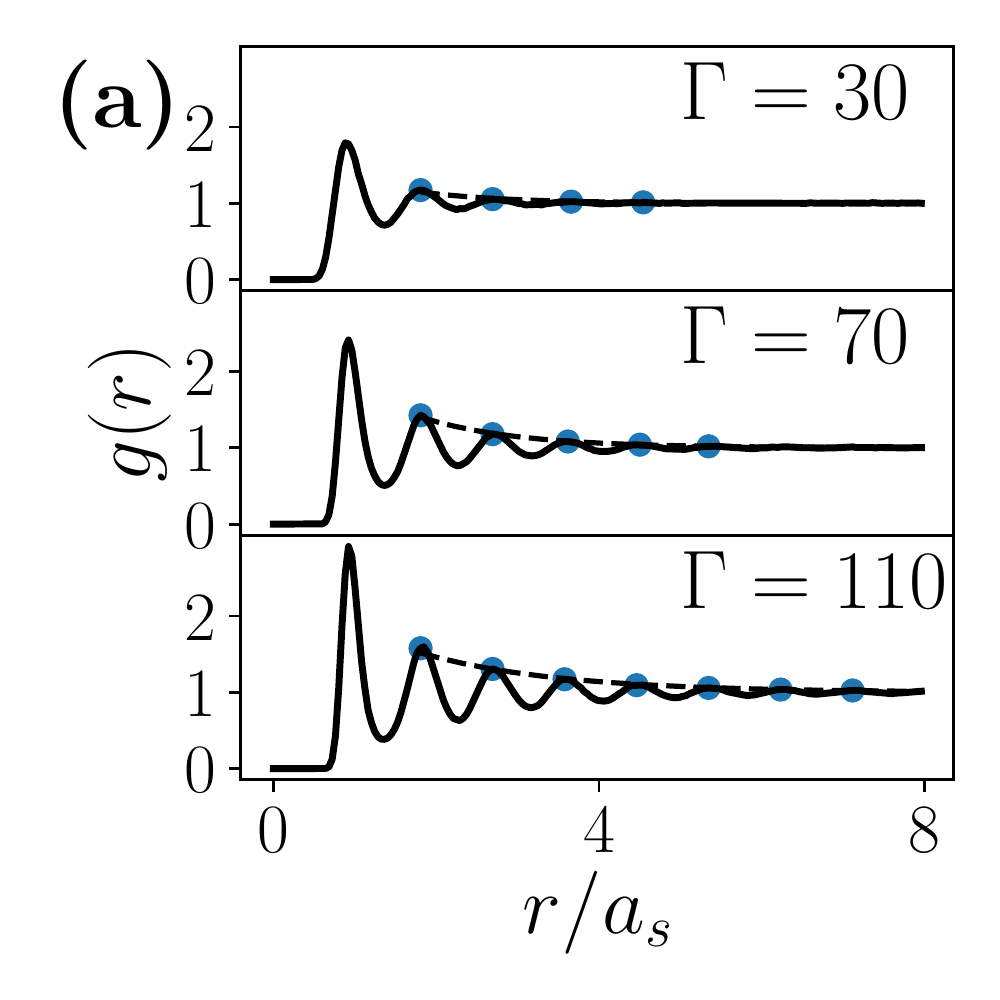}\hfill 
	\includegraphics[width=0.23\textwidth]{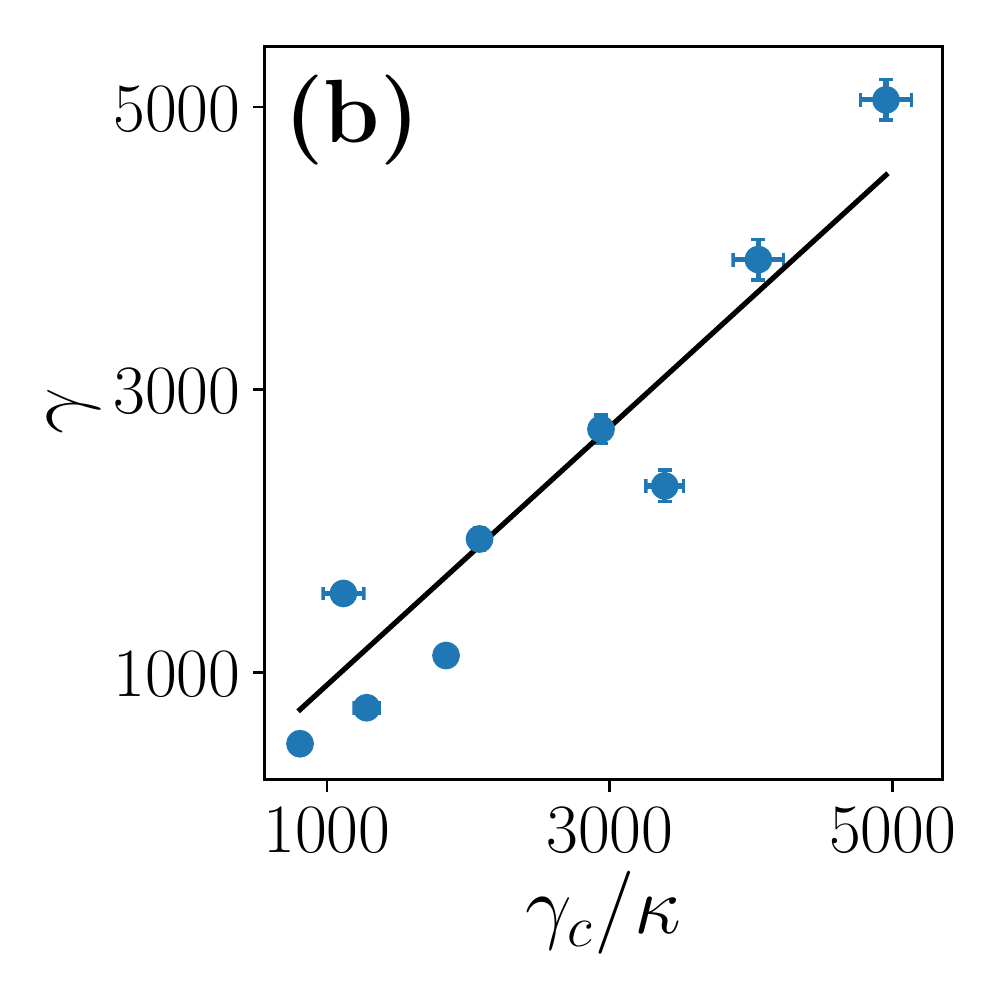}
	\caption{(a) Solid lines: the radial distribution function of the electron liquid in the absence of a periodic potential, Eq.~(\ref{eq:radial_distribution}), for  several values of $\Gamma$. Dashed lines: the exponential fit to the envelope of the oscillations of $g(r)$. (b) Data points: the parameter $\gamma$ in the activation law (\ref{eq:correlated_mobility}). It is extracted for several values of $\Gamma$ from the falloff of $\mu_\perp$ with the increasing potential amplitude using Eq.~(\ref{eq:gamma}). This parameter is plotted against the parameter $\gamma_c$  obtained from the decay of the radial distribution function seen in panel (a). The straight solid line that goes through the origin shows the least squares fit to the relation $\gamma=\gamma_c$. From this fit we estimate $\kappa=0.91\pm0.06$. The error bars represent the errors from approximating the decay of the correlation functions by exponentials using the procedure shown in (a)}
	\label{fig:gamma_vs_xi}
\end{figure}

The radial distribution function is shown in Fig.~\ref{fig:gamma_vs_xi}(a).  It has a familiar for a liquid form of decaying oscillations. The decay is close to exponential. We found the exponent of the decay $\xi$ for different values of $\Gamma$  from the envelopes of the oscillations of $g(r)$, which are shown by the dashed lines in Fig.~\ref{fig:gamma_vs_xi}(a). Except for very large $\Gamma$, the values of $\xi$ are close to the mean inter-electron distance $a_s$. Meanwhile, in Eq.~(\ref{eq:correlated_mobility}) the correlation length was assumed to be large compared to $a_s$. One may, empirically, extrapolate the estimate of $\gamma$ in Eq.~(\ref{eq:correlated_mobility}) to the range $\xi\sim a_s$ by replacing $\xi \to \xi +a_s$, i.e., by setting $\gamma=\gamma_c$, with
\begin{align}
\label{eq:gamma}
\gamma_c = 2\pi\kappa\Gamma (\xi + a_s)^2/a_s^2.
\end{align}
Here, $\kappa$ is a numerical coefficient, $\kappa \sim 1$.

Figure~\ref{fig:gamma_vs_xi}(b) shows a comparison between the factor $\gamma_c$, which is obtained from the pair correlation function of a spatially uniform electron liquid, with the factor 
\begin{align}
\label{eq:gamma_parameter}
\gamma =- \frac{d\ln\mu_\perp}{d(A/E_C)},
\end{align}
extracted from the entirely different simulations of the transverse mobility $\mu_\perp$ in a periodic potential. The proportionality of $\gamma$ and $\gamma_c$ in a broad parameter range  provides a reasonably strong argument in favor of the proposed qualitative picture of the thermally activated mobility in a commensurate periodic potential. Moreover, the results suggest that the correlation length in the electron liquid can be determined by measuring the transverse mobility in a maximally commensurate weak periodic potential.

\section{Transverse mobility for different values of the commensurability parameter}

The previous section described the mobility $\mu_\perp$ where the electron system is a liquid, but the electron crystal with the same density is maximally commensurate with the external potential, the commensurability parameter (\ref{eq:p_parameter}) was $p_c=1$. The electron density in this case was strongly modulated by even a week potential that we used. One can expect that the electron density will be also modulated for $p_c>1$, with the modulation becoming the strongest where the potential is commensurate, including $p_c=\sqrt{3}$ and $p_c=2$, see Appendix \ref{sec:Gaussian_confinement} and \ref{sec:noninteger_pc}. This should reduce the mobility $\mu_\perp$.

For electrons on helium, of interest are the values of $p_c$, which are not large. Indeed, the potential created by the electrodes in Fig.~\ref{fig:electrodes}, exponentially falls off with the decreasing period $2\pi/Q$, for a given height of the helium layer. Therefore the results below are limited to $p_c\leq 2$.

\begin{figure}[h]
	\centering
	\includegraphics[width=0.3\textwidth]{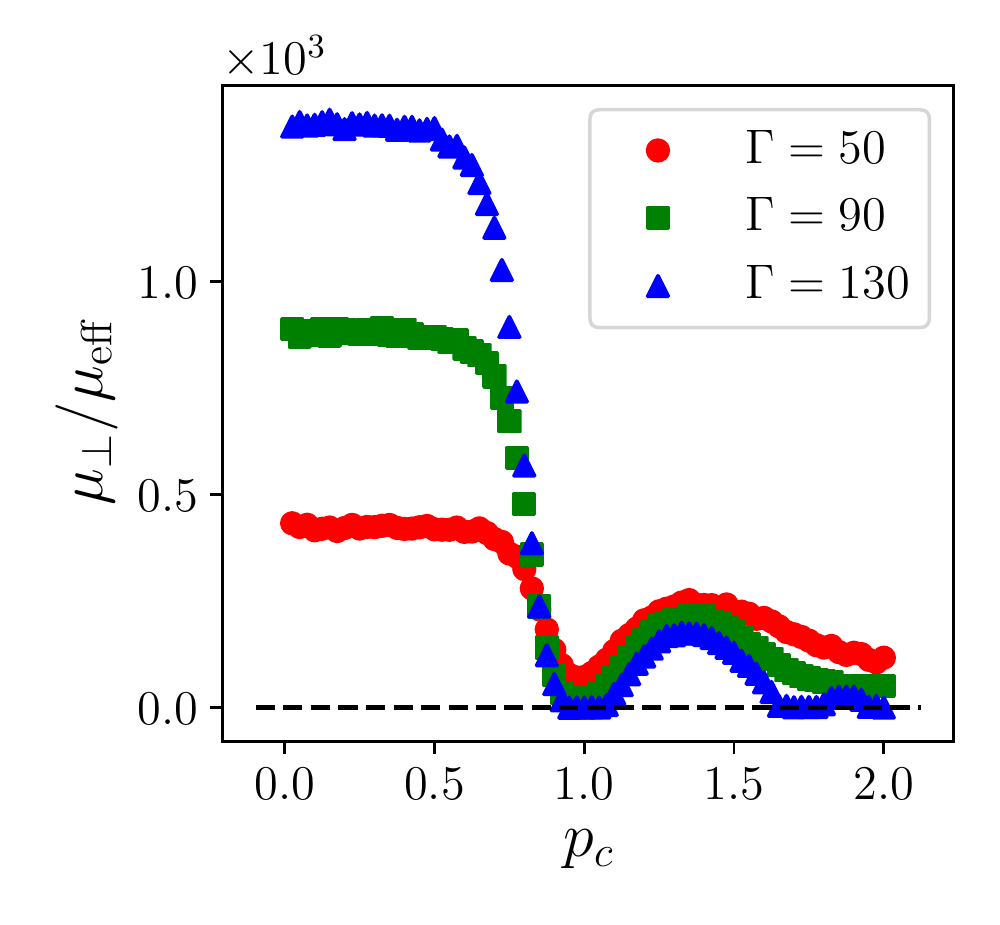}  
	\caption{The mobility $\mu_{\perp}$ of the strongly correlated electron system as a function of the commensurability ratio $p_c$ for $\Gamma = 50$, $90$, and $130$. The scaled potential amplitude us $A/E_C =  0.00157$ for all three cases.}
	\label{fig:mu_vs_p}
\end{figure}

In Fig.~\ref{fig:mu_vs_p}, we show how $\mu_\perp$ varies with the varying $p_c$. In the simulations, $p_c$ is changed by changing the period of the potential, while keeping the mean inter-electron distance $a_s$ fixed. Our simulation cell is a rectangle that contains 1600 electrons and has the dimensions $L_y=40 a_s$ and $L_x=40(\sqrt{3}/2) a_s$. To satisfy the periodic boundary conditions, the period of the potential should be a simple fraction of $L_x$, namely $L_x/n$ with an integer $n$. Keeping in mind that, if the electrons crystallized, there would be 40 electron rows in the $x$-direction, we chose the commensurability parameter to take on the values $n/40$. We used an integer $n$ from $1$ to $80$.

It is seen from Fig.~\ref{fig:mu_vs_p} that a small  periodic potential weakly affects the mobility for $p_c\ll 1$. This is corroborated by the dependence of $\mu_\perp$ on the potential amplitude  $A$ shown in Fig.~\ref{fig:mu_incommensurate} for $p_c = 3/10$.  It is also seen in Fig.~\ref{fig:mu_vs_p} that for all $p_c \lesssim 0.5$ the mobility is close to that in the absence of a periodic potential in the studied range of $A/E_C$. The difference between the values of  $\mu_\perp$ for small $p_c$ and different $\Gamma$ comes from the increase of the rate of scattering by ripplons with the increasing temperature, i.e., with the decreasing $\Gamma$ for the given electron density.

As $p_c$ approaches 1 from below, the mobility sharply drops. The decrease of $\mu_\perp$ becomes steeper with the increasing $\Gamma$. For $p_c$ close to one, the strong electron correlations do not ``compete'' with the periodic potential, electrons can reasonably well fit into the repeated potential minima (perfectly fit, for $p_c=1$). Essentially, the potential competes only with the disorder in the electron liquid. This is why it makes a strong effect even for $A/E_c\sim 10^{-2}$, provided $\Gamma=E_C/k_BT\gg 1$. 

For $\Gamma=130$ and $p_c=1$ the electrons form a Wigner crystal and the mobility becomes zero. For this $\Gamma$, the electrons crystallize also for other commensurate values $p_c=\sqrt{3}$ and $p_c=2$.

In the range $1\lesssim p_c\lesssim 2$, the mobility in the liquid phase is smaller than for $p_c\ll 1$ even for an incommensurate potential. As seen from Fig.~\ref{fig:mu_vs_p}, the dependence on $\Gamma$ is inverted compared to the case of small $p_c$: the larger $\Gamma$, the smaller the mobility. Such behavior suggests that the mobility is limited by the momentum transfer to the potential rather than to the ripplons.

As electron clusters move and rearrange in the incommensurate potential, some electrons go ``uphill'' whereas the others go ``downhill''. The energy gains and losses are not exactly locally balanced, which should lead to an exponential  dependence of $\mu_\perp$ on $A$. Such a dependence has been indeed seen in the simulations, and for $p_c=1.4$ it is presented in Appendix~\ref{sec:noninteger_pc}. The rate of the decrease of $\log\mu_\perp$ with the increasing $A$ is reduced for an incommensurate potential compared to the case $p_c=1$. Also, as Fig.~\ref{fig:mu_vs_p} shows, we did not find a pronounced dip in the mobility near $p_c=\sqrt{3}$ for $\Gamma\lesssim 90$.

It should be noted that the electron vibrations about their quasiequilibrium positions in the electron liquid do not average out the potential. Indeed, it follows from the estimate in Sec.~\ref{subsec:commensurate} that the frequency of electron vibrations is $\sim \omega_p$ for $A/E_C\ll 1$ and $p_c\sim 1$. Then  the mean-square displacement of an electron about its quasiequilibrium position $(k_BT/m\omega_p^2)\sim a_s^2/\Gamma$ is small compared to the squared period of the potential for $p_c\sim 1$ and $\Gamma\gg 1$, cf. Appendix~\ref{sec:Gaussian_confinement}.

\section{Conclusions}

The presented numerical simulations of the  electron liquid on helium have shown several effects of the strong electron correlations on the electron mobility. The results suggest a way to directly characterize these correlations in an experiment.  

For a spatially uniform electron system, the simulations have confirmed the mechanism of transport inferred in the earlier work \cite{Dykman1979a,Buntar'1987,Dykman1993b}. The underlying picture in that work is that the electron scattering by the helium excitations is weak, whereas the electron-electron interaction is strong. Yet in the classical regime $k_BT\gg \hbar\omega_p$ and in the absence of a magnetic field, the major effect of the electron-electron interaction on the mobility is the inter-electron momentum exchange, which is faster than the momentum exchange with the helium excitations. This picture  allowed calculating the mobility explicitly. The excellent agreement of our simulations with the analytical results provides a quantitative basis for the above picture.  Moreover, the simulations have explicitly shown that, when the electron system is slightly heated by the external field, the distribution of the electron kinetic energy is of the Maxwell-Boltzmann form in the co-moving frame, with a temperature higher than the temperature of the helium.

The central results of the paper refer to the mobility of the electron liquid placed into a sinusoidal one-dimensional potential. We studied the parameter range where the potential amplitude $A$ was two orders of magnitude smaller than the electron correlation energy $E_C$. Yet the effect of the potential can be strong. It depends on the interrelation between the mean inter-electron spacing $a_s$ and the period of the potential $2\pi/Q$. More specifically, it depends on whether the potential is commensurate with the electron crystal with the same spacing, even though the results refer to the region where the electrons form a liquid, not a crystal. 

We found that, for $a_sQ/2\pi\ll 1$, the effect of an incommensurate potential on the mobility is effectively washed out by strong electron correlations. Even where $A/k_BT \gtrsim 1$ and the single-electron mobility is strongly reduced, the many-electron mobility shows a very small change. This can be understood by noticing that, for $E_C\gg A$, the electron system averages out the potential. At the same time, the potential is smooth on the electron thermal wavelength, and therefore the potential does not add to the electron scattering.

A qualitatively different behavior is displayed in a commensurate potential. Even where the two-particle correlation function is weakly modified compared to the case of no potential, the electron density becomes periodically modulated with an amplitude significantly larger than in the single-electron picture. The mobility transverse to the potential troughs $\mu_\perp$ displays an exponential dependence on the potential amplitude $A$. We associate the very mechanism of the mobility with the absence of long-range order. In the picture suggested by the results, the mobility results from the correlated many-electron activated transitions within areas with the typical size given by the correlation length in the liquid. 

We have found a simple relation between the correlation length in the electron system in the absence of the periodic potential and the activated fall-off of the mobility with the increasing $A$ for $A\ll E_C$. This relation provides a means for measuring the electron correlation length in the experiment. We are not aware of other means to measure the correlation length of the nondegenerate electron liquid on the surface of helium.

The parameters used in the simulations are within the typical range of the parameters used in the experiments on electrons on helium, and the proposed weak one-dimensional potential with a period $\sim 1~\mu$m can be implemented with conventional technology. The system is advantageous for studying the effect of commensurability given that the electron density can be easily varied. Our results show that these effects are strong even for a classical electron liquid. As a future direction, it would be interesting to study these effects where the electron dynamics becomes quantum, even though the system remains nondegenerate and therefore there is no need to have high electron densities, which significantly simplifies an experimental implementation. 

\acknowledgments
This research of KM was supported in part by the National Science Foundation, Grant DMR 1708331. MID was supported in part by  the Grant DE-SC0020136 funded by the  US Department of Energy, Office of Science.

\appendix

\section{Outline of the simulations}
\label{sec:simulations}

\subsection{Matrix elements of coupling to the helium excitations}

A key distinctive part of the simulations is the incorporation of the actual elastic and inelastic scattering into the electron equations of motion. The scattering is due to the coupling (\ref{eq:coupling}) to ripplons and phonons in liquid helium.  The coupling parameters $V_{\qb\,\alpha}$ are well-known \cite{Andrei1997,Dykman2003a,Schuster2010,Moskovtsev2019}. We give them here for completeness. 

We are interested in the case of a weak field that presses electrons to the surface. Then the coupling to ripplons comes primarily from the ripplon-induced change of the image potential that attracts the electrons to the helium surface. The image potential is also changed by phonons, as they modulate the helium density and thus the dielectric constant. The corresponding changes of the potential energy of an electron at a distance $z$ from the surface are, respectively,
\begin{align}
\label{eq:matrix_elements}
&V_{\qb}^{\mathrm {(rp)}}(z)=\Lambda\frac{(\hbar q)^{1/2}}{(2\rho\omega_{\qb}S)^{1/2}}\left\{z^{-2}[1-qz K_1(qz)]\right\}_z,\nonumber\\
&V_{\qb,q_z}^{\mathrm {(ph)}}(z)=-i\Lambda q(\hbar\omega_{\qb,q_z}/V v_{\rm He}^2\rho_{\rm He})^{1/2}\nonumber\\
&\qquad \times \int_{-\infty}^0 dz'\frac{\sin (q_z z')}{z-z'}K_1\bigl(q(z-z')\bigr).
\end{align}
Here $S$ is the area of the system, $\Lambda = e^2(\epsilon -1)/4(\epsilon+1)$ ($\epsilon$ is the dielectric constant of helium, $\epsilon\approx 1.057$), and $K_1(x)$ is the Bessel function. The parameters $V_{\qb,\alpha}$ of the coupling Hamiltonian  (\ref{eq:coupling}) are obtained by calculating the diagonal matrix elements of the functions (\ref{eq:matrix_elements}) on the wave function $\psi_0(z)$ of the ground state of electron motion normal to the helium surface.

\subsection{The calculation}
\label{subsec:calculation}
The simulation procedure used in this work is described in detail in \cite{Moskovtsev2019}. Here we briefly summarize the most important aspects. The simulations were conducted using HOOMD-Blue \cite{Anderson2008,Glaser2015} with a custom interaction potential, integrator, and external forces.

As indicated in the main text, we consider $N=1600$ electrons placed into a rectangular cell with the aspect ratio of $L_x/L_y = \sqrt{3}/2$ and periodic boundary conditions. If an electron crosses the cell boundary, it is introduced back into the cell from the opposite side. Electrons interact via a long-range Coulomb force which is handled through the Ewald summation \cite{Gann1979}. We consider an  external potential $\eqref{eq:Hamiltonian_ee}$ imposed on electrons, which is periodic along the $x$-axis. The period is chosen so that the wave vector $Q$ satisfies the condition $QL_x = 2\pi n$ with a positive integer $n$. The time step is chosen to be $\Delta t \approx (2\pi/\omega_p)/50$. We have checked that the results do not change upon reducing the step further.

We integrate equations of motion with the velocity Verlet algorithm modified for direct simulation of scattering events by excitations in helium. The inelastic scattering provides a thermal bath that maintains a set temperature in the electron system. However, since the inelastic scattering rate is low, the effective temperature of the system may deviate slightly from the bath temperature. We define the effective temperature in a non-equilibrium state as follows:
\begin{equation}
T_{\mathrm{eff}}(t) \equiv (m/2Nk_B)\sum_n (\vb_n(t) - \vb_d)^2,
\end{equation}
where
\begin{equation}
\vb_d \equiv \frac{1}{NN_t}\sum_{n=1}^N\sum_{s=0}^{N_t} \vb_n(s\Delta t)
\end{equation}
is the drift velocity of the system averaged over time. $N_t$ is the number of time steps and $s$ is the time step index.

In a typical simulation, we prepare the system in a triangular lattice with one side of the triangle oriented along the $y$ axis. Initial velocities are randomly oriented with the absolute velocity values distributed according to the Boltzmann distribution for a set temperature. External drive electric field $E_d$ is applied along the $x$ axis. We allow the driven system to stabilize in a stationary state for $10^6$ time steps before collecting data. This stabilization time is sufficient for the relatively high temperatures considered here.

We found that the results did not change when the system size changed from $1024$ to $1600$ electrons. Some results were also checked for different aspect ratios of the simulation box. The absence of changes indicates that the system size effects are within the simulation uncertainty. 

The natural parameter of the state of the electron system in the classical regime is $\Gamma =e^2(\pi n_s)^{1/2}/k_BT$. However, the electron mobility depends on $T$ separately. In our finite system we cannot vary the density $n_s$ continuously while maintaining a given ratio of the inter-electron distance and the periodic potential. Therefore our results for different $\Gamma$ are obtained by varying the temperature.

\section{Electron density in the maximally commensurate potential}
\label{sec:Gaussian_confinement}

\begin{figure}[h]
	\includegraphics[scale=0.25]{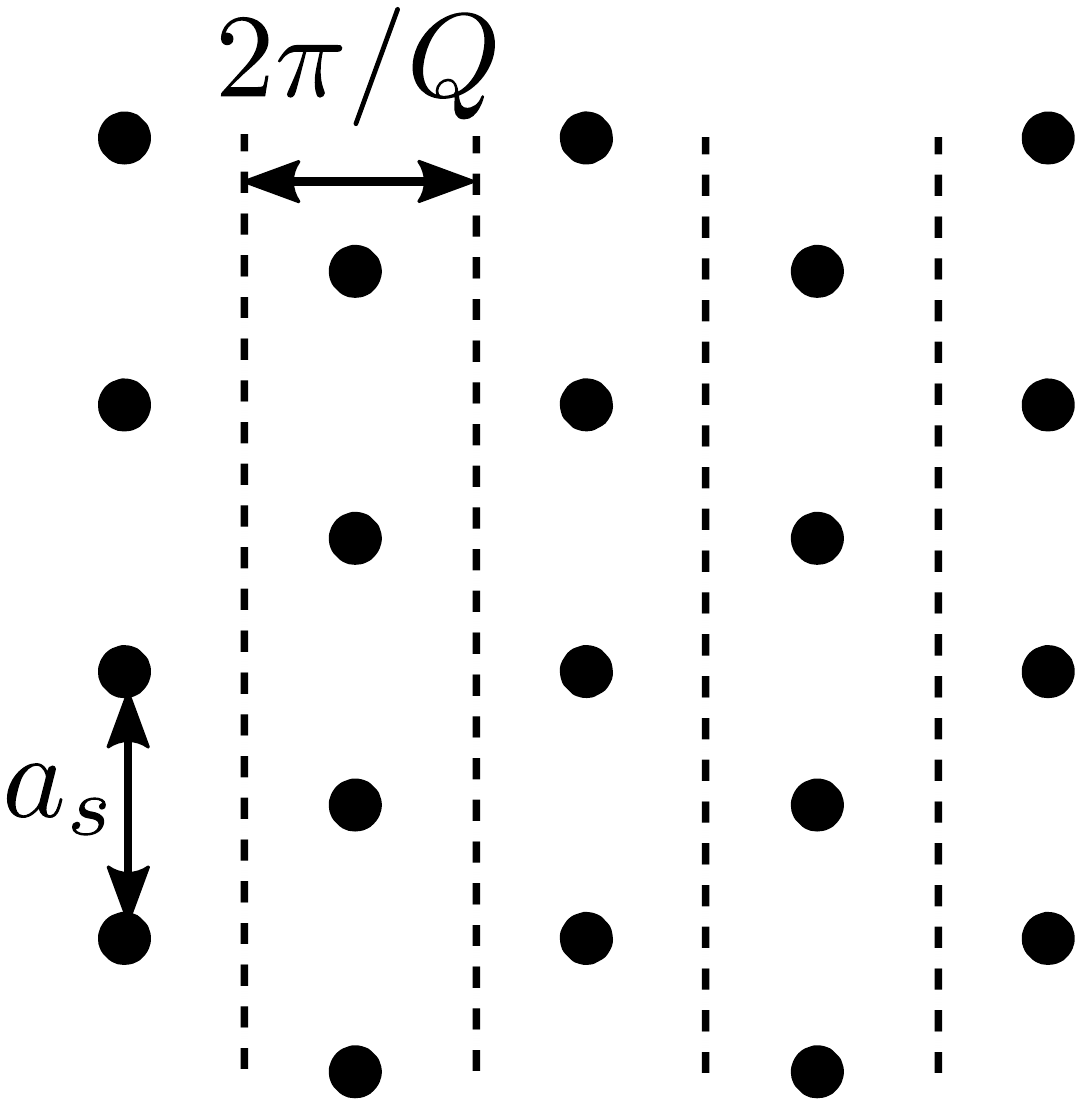}~\quad\quad\quad\quad
	\includegraphics[scale=0.25]{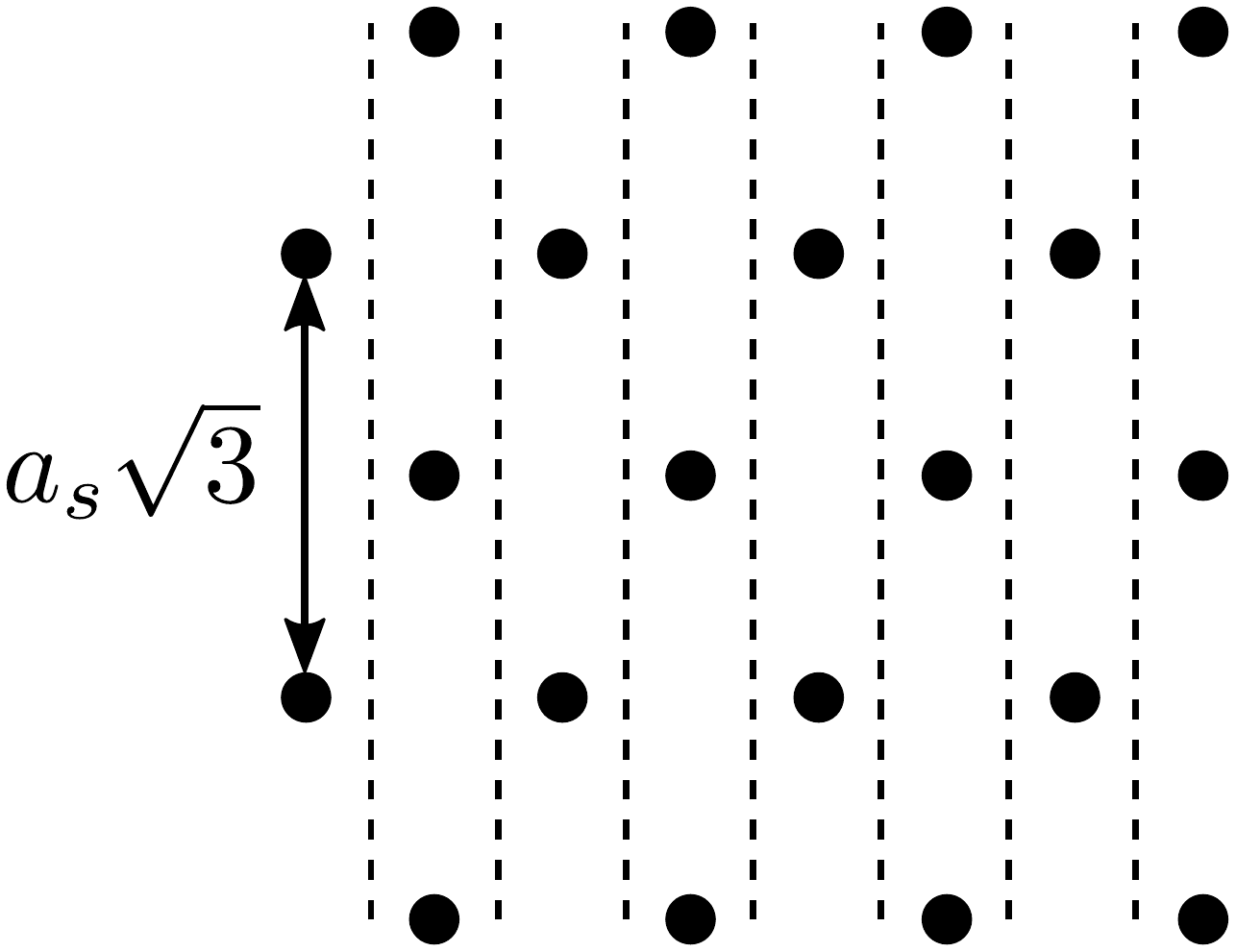}
	\caption{Schematic of a triangular lattice placed in a 1D periodic potential $U=-A\cos Qx$. The dashed lines show the maxima of the potential. The left panel refers to what we call the maximally commensurate potential, $p_c = 1$, whereas the right panel refers to $p_c=\sqrt{3}$. This sketch illustrates the definition~(\ref{eq:p_parameter}) of the commensurability parameter $p_c$. In simulations we study the liquid phase with no crystalline order.}
	\label{fig:commensurability_scheme}
\end{figure}

Two simple types of the commensurate potentials are shown in Fig.~\ref{fig:commensurability_scheme}. For the potential shown in the left panel, the distance between the electrons in the same potential trough is equal to the mean interelectron distance $a_s$ and $p_c=1$. For the potential shown in the right panel, this distance is $a_s\sqrt{3}$ and $p_c=\sqrt{3}$. Respectively, the distances between the troughs differ by the factor $\sqrt{3}$. Each of these potentials leads to a series of potentials with $p_c=n$ or $p_c= n\sqrt{3}$, into which the electron crystal can fit with no distortion. As mentioned in the text, we call the potential with $p_c=1$ the maximally commensurate potential.

If the electrons  are placed into a sufficiently strong maximally commensurate periodic potential ($p_c=1$) but do not crystallize, one can approximate the density as uniform along the potential troughs and a sum of Gaussian peaks in the transverse direction,
\begin{align}
\label{eq:Gauss_density}
\rho_s(\rb) &\approx \frac{n_s}{\varkappa Q}\sqrt{2\pi}\sum_n\exp[-(x-2\pi n/Q)^2/2\varkappa^2].
\end{align}
The distribution  $\rho(\rb)$ corresponds to a ``single-electron'' wire: one electron per trough with the mean inter-electron distance along the trough equal to the mean inter-electron spacing $a_s$.
The width of the peaks $\varkappa$ can be roughly estimated by assuming Boltzmann distribution about the minima of $U(\rb)$ and further assuming that the potential near a given minimum, in addition to $U(\rb)$, has a contribution from the electron ``wires'' localized in other minima, i.e., disregarding the short-range ordering along the wire. This gives $\varkappa = [k_BT/(AQ^2 + \frac{1}{3}\pi e^2 n_s Q)]^{1/2}$. For the considered case of maximum commensurability this can be also written as $\varkappa=(k_BT/Q^2)^{1/2}[A +E_C(\sqrt{3}/72\pi)^{1/2}]^{-1/2}$. According to this estimate, the major factor in the width of the peaks is the electron-electron interaction, for the considered range of the potential strength $A\ll E_C$. 

From Eq.~(\ref{eq:Gauss_density}), for a strongly confined liquid, the Fourier components $\alpha_m$ of the electron density in Eq.~(\ref{eq:density_periodic}) are 
\begin{align}
\label{eq:Fourier_Gauss}
\alpha_m=2 \exp(-m^2\varkappa^2Q^2/2).
\end{align}
They fall off exponentially fast with the increasing $m$ for $m\gg 1/\varkappa Q$. 

\section{Electron correlations and the mobility for $p_c>1$}
\label{sec:noninteger_pc}
We defined $p_c$ using the Wigner crystal with reciprocal lattice vectors ${\bf G}_1 = 2\pi a_s^{-1}(-1/\sqrt{3}, 1)$ and ${\bf G}_2 = 2\pi a_s^{-1}(2/\sqrt{3},0)$, so that the maximal commensurability $p_c=1$  corresponded to the reciprocal lattice vector of the one-dimensional periodic potential $Q=G_{2x}$. Other commensurate values of $Q$ for a one-dimensional potential are ${\bf Q}=n_1(2{\bf G}_1 +{\bf G}_2) + n_2 {\bf G}_2$ with integer $n_1, n_2$. Respectively, the smallest values of $p_c$ that correspond to a commensurate lattice are $p_c= 1, \sqrt{3}, 2,...$. 
\begin{figure}[t]
	\includegraphics[scale=0.37]{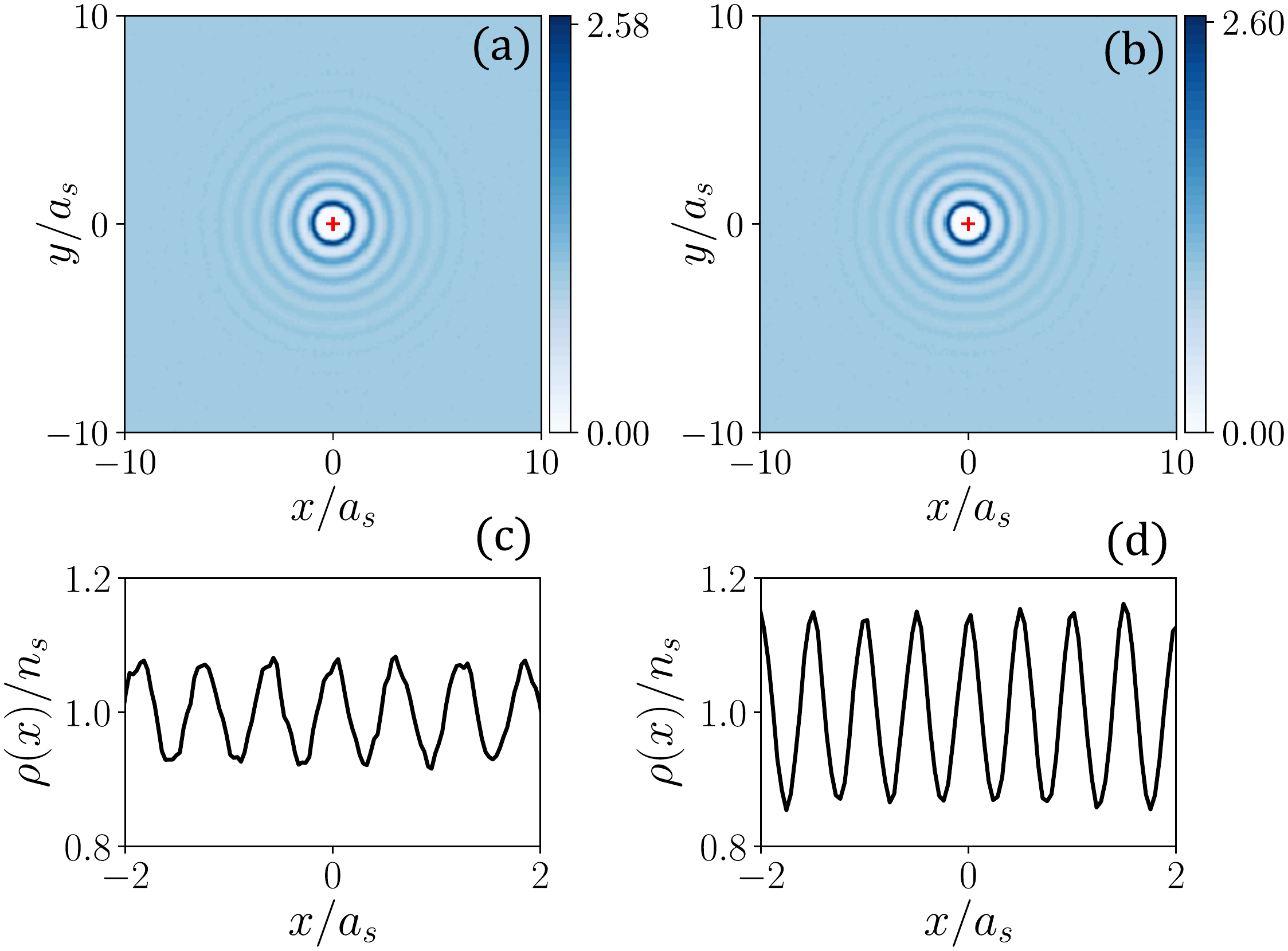}
	\caption{Plots (a) and (b) show the pair correlation function of the electron liquid $\tilde{g}^{(2)}(\rb)$, Eq.~(\ref{eq:pair_correlation}), for $p_c=1.4$  and  $p_c=1.725\approx \sqrt{3}$ respectively. Plots (c) and (d) show the averaged over time electron density along the $x$ direction for the same parameters as in (a) and (b), respectively. The scaled amplitude of the potential is $A/E_C=0.00157$, and the plasma parameter is $\Gamma=90$ in all panels.}
	\label{fig:pair_distribution_noninteger_pc}
\end{figure}

As shown in the main text, in the range $1\lesssim p_c \lesssim 2$ the transverse mobility $\mu_\perp$ is significantly smaller than in the absence of the periodic potential. This is related to the partial adjustment of the electron liquid to the potential. To demonstrate this adjustment, in Fig.~\ref{fig:pair_distribution_noninteger_pc} we show the density modulation and the pair correlation function for   $p_c=1.4$, which lies between the two smallest values corresponding to the commensurability, and for $p_c=1.725$, which is close to the commensurate case $p_c=\sqrt{3}$.

\begin{figure}[h]
\includegraphics[scale=0.45]{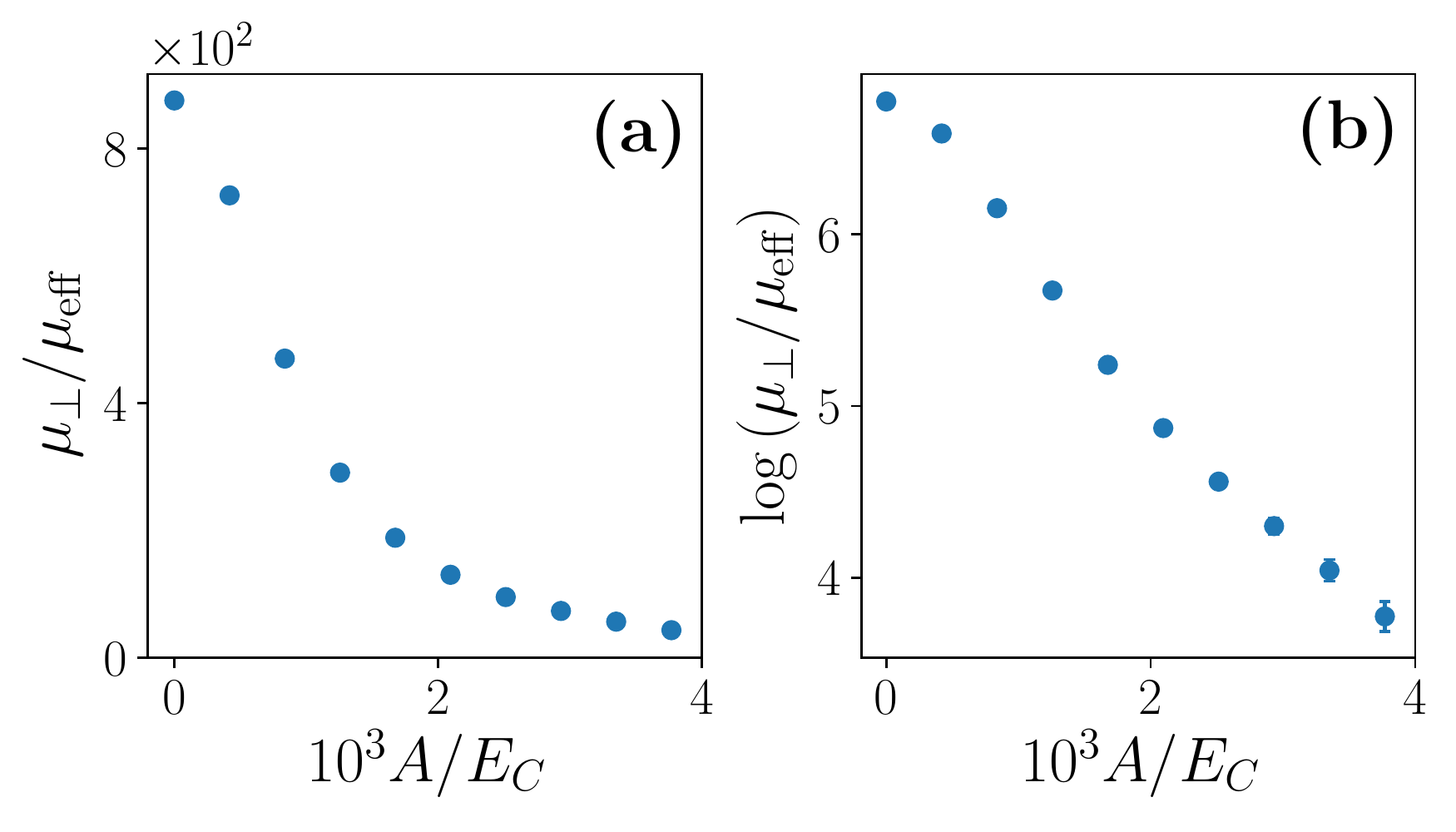}
\caption{The mobility transverse to the potential troughs as a function of the potential amplitude $A$ for $p_c = 1.4$ on the linear (a) and logarithmic (b) scales. The plasma parameter $\Gamma = 90$.}
\label{fig:mu_vs_A_p_1p4}
\end{figure}

It is seen from Figs.~\ref{fig:pair_distribution} and \ref{fig:pair_distribution_noninteger_pc} that the pair correlation function weakly depends on $p_c$ for the considered weak periodic potential. In contrast, the density modulation is significantly stronger in the commensurate case than away from commensurability. Moreover, it is significantly stronger for $p_c=1$ than for $p_c=\sqrt{3}$, i.e., for the electrons being spaced more closely within the potential trough and, respectively, for a larger distance between the troughs, cf. Fig.~\ref{fig:commensurability_scheme}. 

Fig.~\ref{fig:mu_vs_A_p_1p4} shows that the electron mobility displays an exponential dependence on the potential amplitude even away from the commensurability, where $p_c=1.4$. However, the values of the mobility are much larger than in the maximally commensurate case $p_c=1$ shown in Fig.~\ref{fig:no_ripplons}. This is a consequence of the partial averaging of the incommensurate potential in the electron liquid.



\section{Many-electron relaxation time in the presence of inelastic scattering}
\label{sec:Inelastic}

The goal of this Appendix is to derive the quantum kinetic equation for the strongly correlated electron system on helium with the account taken of inelastic scattering. We will follow the steps outlined in the analysis of elastic scattering \cite{Dykman1997}. At the heart of the analysis is the assumption that the rate at which the electrons exchange energy and momentum with each other, which is characterized by the plasma frequency $\omega_p=(2\pi e^2 n_s^{3/2}/m)^{1/2}$ ($n_s$ is the electron density), is much faster than the rate of scattering by the excitations in helium.   

To simplify the analysis, we will assume that there is no magnetic field applied to the electron system. Incorporating the field is straightforward, similar to how it was done for the case of elastic scattering. We will further assume that the electron system is nearly  classical, 
\begin{equation}
\label{eq:classical}
k_BT \gg \hbar\omega_p \sim e\Ef\lambdabar
\end{equation}
where $\lambdabar=\hbar/(2mk_BT)^{1/2}$ is the thermal wavelength and $\Ef$ is the fluctuational electric field that drives an electron due to the density fluctuations in the electron system. We note that,  in the case of elastic scattering, quantum corrections to the scattering rate $\sim (e\Ef\lambdabar/k_BT)^2/48$ contains a small numerical factor, which suggests that the theory applies even where the ratio $k_BT/\hbar\omega_p$ is not very small. 

When the condition (\ref{eq:classical}) holds, an electron has
a well-defined kinetic energy $p^2/2m \sim k_BT$ and a well-defined
potential energy in the field of other electrons. The uncertainty of each
of these energies is determined by the smearing of the electron wave
packet $\lambda$\raisebox{.6ex}{\hspace{-11pt}--}\hspace{4pt}. For
an electron in an electric field ${\bf E}_{\rm fl}$ this uncertainty is
given by $e\Ef\lambdabar$, and it is
small compared to $k_BT$. 

\subsection{Transport equation in the operator form}

The long-wavelength  many-electron conductivity $\sigma_{xx}(\omega)$  is expressed in the standard way, using the Kubo formula, in terms of the Fourier transform of the correlator of the total electron momentum  $\hat \Pb = \sum_n\hat\pb_n$,
\begin{align}
\label{eq:P_correlator_general}
&\langle \hat P_x (t)\hat P_x(0)\rangle = 
{\rm Tr}_e\left[e^{i \hat H_{ee}t} \hat P_xe^{-i\hat H_{ee}t}\hat{\cal G}_x(t)\right],\nonumber\\
&\hat{\cal G}_x(t) = Z^{-1}{\rm Tr}_{\rm He} 
\left[\hat S(t)\hat P_xe^{-\beta \hat H}\hat S^+(t)\right],\; \nonumber\\
&\hat S(t) = \exp(i \hat H_{ee}t)\exp(-i\hat Ht)\quad (\hbar =1).
\end{align}
Here and below we set $\hbar =1$; Tr$_{e}$ and Tr$_{\rm He}$ are traces over the wave functions of the
isolated electron system and of the helium vibrations, and $Z = {\rm Tr}_e{\rm Tr}_{\rm He}
\exp(-\beta \hat H)$ is the partition function ($\beta \equiv 1/k_BT$); $\rb_n$ and $\pb_n$ are the 2D the coordinate and momentum of the $n$th electron.

The operator $\hat{\cal G}_x(t)$ is the density matrix of the many-electron system. To the lowest order in the coupling to helium excitation, the many-electron transport equation for $\hat{\cal G}_x$ can be  written in the operator form as
\begin{align}
\label{eq:operator_QKE}
&{\partial \hat{\cal G}_x(t) \over \partial t} = -{\rm Tr}_{\rm He}\int_0^t
dt'\left[\hat H_i(t),\left[\hat H_i(t'), \hat \rho_{\rm He}\hat{\cal G}_x(t)\right]\right]\nonumber\\
&\hat H_i(t) = \exp[i(\hat H_{ee}+ \hat H_{\rm He})t]\hat H_i
\exp[-i( \hat H_{ee}+ \hat H_{\rm He})t];\nonumber\\
& 
\hat \rho_{\rm He} =\exp(-\beta \hat H_{\rm He})/{\rm Tr_{He}}\exp(-\beta \hat H_{\rm He})
\end{align}
Here 
\begin{align}
\label{eq:coupling_SM}
\hat H_i = \sum_n\sum_{\qb,\alpha}V_{\qb\,\alpha}e^{i\qb\hat\rb_n}(\hat a_{\qb\,\alpha}+\hat a_{-\qb\,\alpha}^\dagger)
\end{align}
is the coupling Hamiltonian; $\hat a_{\qb\,\alpha}$ is the annihilation operator of the vibrational mode in helium with quantum numbers $\qb,\alpha$ ($\qb$ is the 2D wave vector of the mode).  

In deriving Eq.~(\ref{eq:operator_QKE}) we assumed that $t \gg t_{\rm coll},\hbar (k_BT)^{-1}$. The quantity $t_{\rm coll}$ is the characteristic duration of a collision of an electron with a helium excitation. It gives  the width of the interval $t-t'$ that contributes to the integral over $t'$. This interval is supposed to be small compared to  the relaxation time $\tau$ over which $\hat{\cal G}_x(t)$ varies. The approximation (\ref{eq:operator_QKE}) corresponds to the ladder approximation in the single-electron transport theory. It takes into account the interaction energy $\hat H_i$ multiplied by a long time $t\sim \tau$ while the term $\beta \hat H_i$ is disregarded in the considered range of comparatively high temperatures. 

It is essential that the scattering by ripplons and phonons is short-range: the density of states of ripplons and phonons increases with the increasing wave number $|\qb|$, and the values of $|\qb|$ are essentially limited by the reciprocal size of the electron wave package $\sim\lambdabar^{-1}$. In a strongly correlated electron system at most one electron at a time can collide with a short-range scatterer. Therefore short-range scattering can be described in the ``single-site'' approximation, cf.~\cite{Dykman1982b}. In this approximation only diagonal terms are retained in the double sum over the electrons that enters the product $\hat H_i(t)\hat H_i(t')$. We can then write
Eq.~(\ref{eq:operator_QKE}) as
\begin{widetext}
\begin{align}
\label{eq:QKE_operator_explicit}
&{\partial \hat{\cal G}_x \over \partial t} = -\sum_{\qb,\alpha}
\left|V_{\qb\,\alpha}\right|^2\sum_n\int_0^t
dt' \left[e^{i\qb\hat\rb_n(t)} e^{-i\qb\hat\rb_n(t')}\phi_{\qb\,\alpha}(t-t')\hat{\cal G}_x(t)
+\hat{\cal G}_x(t)  e^{i\qb\hat\rb_n(t')} e^{-i\qb\hat\rb_n(t)}\phi_{\qb\,\alpha}(t'-t)\right.\nonumber\\
&\left.- e^{i\qb\hat\rb_n(t)}\hat{\cal G}_x(t) e^{-i\qb\hat\rb_n(t')} \phi_{\qb\,\alpha}(t'-t)
 -e^{i\qb\hat\rb_n(t')}\hat{\cal G}_x(t) e^{-i\qb\hat\rb_n(t)} \phi_{\qb\,\alpha}(t-t')\right]
\end{align}
\end{widetext}
where 
\[\phi_{\qb\,\alpha}(t)=(\bar n_{\qb\,\alpha}+1)\exp(-i\omega_{\qb\,\alpha}t) +\bar n_{\qb\,\alpha}\exp(i\omega_{\qb\,\alpha}t)\] 
is the Green function of the mode $(\qb,\alpha)$ ($\bar n_{\qb\,\alpha}\equiv \bar n(\omega_{\qb\,\alpha})$ is the mode Planck number), and 
\begin{align}
\label{eq:initial_condition}
\hat\rb_n(t) = e^{i\hat H_{ee}t}\hat\rb_ne^{-i\hat H_{ee}t}; \;
\hat{\cal G}_x(0) = Z_{ee}^{-1}\hat P_xe^{-\beta \hat H_{ee}}
\end{align}
with $Z_{ee} = {\rm Tr}_e \exp(-\beta\hat H_{ee})$.

\subsection{The general form of the many-electron density operator}

In the absence of a magnetic field, the operator of the total electron momentum $\hat \Pb$ commutes with $\hat H_{ee}$. Therefore $\hat{\cal G}_x(t)$ is the only time-dependent term in the momentum correlation function  (\ref{eq:P_correlator_general}). We are interested in the diagonal matrix elements of $\hat{\cal G}_x(t)$ on the eigenfunctions of the many-electron Hamiltonian $\hat H_{ee}$. 

Alternatively, and equivalently, instead of $\hat{\cal G}_x(t)$ we could consider  the operator $\exp(-i\hat H_{ee}t)\hat{\cal G}_x(t)\exp(i\hat H_{ee}t)$, which has the same diagonal matrix elements. The off-diagonal matrix elements of this operator decay over the characteristic time of the many-electron dynamics $\omega_p^{-1}$. Over this time the many-electron system comes to thermal equilibrium with respect to a frame that moves with the velocity determined by the initial conditions \cite{Landau1980}; the effective temperature is also determined by the initial conditions. In the considered case this temperature is equal to the temperature of the helium excitations. 

The understanding of the time evolution of the momentum correlator (\ref{eq:P_correlator_general}) relies on two observations. First, there is no energy exchange between the two thermal reservoirs, the vibrational excitations in helium and the many-electron system, as they are both at the same temperature. Second, there is a momentum exchange. Since the vibrational reservoir has much more degrees of freedom than the electron system, the momentum of the electron system decays, and so does the correlator (\ref{eq:P_correlator_general}). 

The above observations show that, with the account taken of the initial condition (\ref{eq:initial_condition}), on the times much larger than $\omega_p^{-1}$ the operator $\hat{\cal G}_x(t)$ can be sought in the form
\begin{align}
\label{eq:g_x_operator}
\hat{\cal G}_x(t) \approx Z_{ee}^{-1}g(t)\hat P_x\exp(-\beta \hat H_{ee}).
\end{align} 
This solution reflects the symmetry of $\hat{\cal G}_x(0)$ as a component of a vector, which is preserved by the coupling to the thermal reservoir, and the fact that $\hat{\cal G}_x(t) $ is diagonal on the eigenfunctions of $\hat H_{ee}$. Since the momenta of different electrons are uncorrelated (see also below), the total momentum $\Pb$ is small, and therefore only a linear term in $\hat \Pb$ is held in Eq.~(\ref{eq:g_x_operator}).  Function $g(t)$ describes the decay of the momentum correlator, with $g(0)=1$.

\subsection{Quantum transport equation in the Wigner representation}

It is convenient to us the Wigner representation to analyze the electron dynamics in the classical regime while taking into account the quantum nature of the electron scattering by helium excitations. We start with the basis states of the many-electron system as plane waves
\begin{equation}
\label{eq:plane_waves}
\left|\{{\bf k}_n\}\right\rangle \equiv\prod_n (2\pi)^{-1}\exp\left(i
{\bf k}_n{\bf r}_n\right).
\end{equation}
A many-electron operator in the Wigner representation has the form 
\begin{align}
\label{eq:Wigner}
&K\left(\{{\bf p}_n\},\{{\bf r}_n\}\right)=  \int \left[\prod_n
d\mbox{\boldmath $\zeta$}_n\exp\left(i \mbox{\boldmath $\zeta$}_n{\bf
r}_n\right)\right]
\nonumber \\
& \times\left\langle\{{\bf p}_n+{1\over 2}\mbox{\boldmath $\zeta$}_n\}\right|
\hat K\left(\{\hat\pb_n\},\{\hat\rb_n\}\right)\left|\{{\bf p}_n-{1\over 2}\mbox{\boldmath $\zeta$}_n\}\right\rangle.
\end{align}
The correlator (\ref{eq:P_correlator_general}) can be written as 
\begin{eqnarray}
\label{eq:P_correlator_Wigner}
&&\langle \hat P_x(t)\hat P_x(0)\rangle  =  
\int\hspace{-4pt}
\int \left[\prod_n (2\pi)^{-2}\,d{\bf p}_n d{\bf r}_n \right] \nonumber \\
&& \hspace{0.2in} \times
P_x\left(\{{\bf p}_n\}\right) 
G_x\left(t;\{{\bf p}_n\},\{{\bf r}_n\}\right),
\end{eqnarray}
where $G_x\left(t;\{{\bf p}_n\},\{{\bf r}_n\}\right)$ is the
matrix element of the operator $\hat{\cal G}_x(t)$ in the Wigner representation and $\Pb\left(\{\pb_n\}\right) = \sum_n \pb_n$.

\begin{widetext}
The equation for $G_x\left(t;\{{\bf p}_n\},\{{\bf r}_n\}\right)$ follows from Eq.~(\ref{eq:QKE_operator_explicit}). The characteristic range of $t'$ that contributes to
the integral over $t'$ was shown to be small ($\lesssim 1/k_BT$) for elastic scattering \cite{Dykman1997}; it is also small for inelastic scattering, as the energy transfer is $\sim k_BT$, and therefore for the both scattering mechanisms $t-t'\lesssim (k_BT)^{-1}$. Over the time $t-t'$, the change of the electron momentum due to the fluctuational electric field $\sim e \Ef (t-t')$ is small compared to the thermal momentum $p_T=(mk_BT)^{1/2}$ in the range (\ref{eq:classical}), $e\Ef(t-t')/p_T \lesssim e\Ef\lambdabar/k_BT \ll 1$. Therefore one  can approximate
\begin{align}
\label{eq:short_time_expansion}
&\hat{\bf r}_n(t') = \hat{\bf r}_n(t) - {1\over m}(t-t')\hat \pb(t),\nonumber \\
&\exp[-i\qb\hat\rb_n(t')] = \exp[-i\qb\hat\rb_n(t)]
 \exp[i\qb\hat\pb_n(t)(t-t')/m]\,\exp[-iq^2(t-t')/2m].
\end{align}
Then the equation for $G_x$ takes the form

\begin{align}
\label{eq:QKE_Wigner}
&{\partial G_x\left(t;\{\pb_n\},\{\rb_n\}\right) \over \partial t} = -\sum_{\qb,\alpha}
\left|V_{\qb\,\alpha}\right|^2\sum_{n'}
\int_0^t dt_1 
\xi(t-t_1; \qb,\,\pb_{n'})\Bigl[\phi_{\qb\,\alpha}(t-t_1)G_x\left(t;\{\pb_n\},\{\rb_n\}\right)
\nonumber\\
& - \phi_{\qb\,\alpha}(t_1-t)G_x\left(t;\{\pb_n-\qb\delta_{nn'}\},\{\rb_n\}\right) \Bigr] + {\rm c.c.}, \qquad \xi(t;\qb,\,\pb) = 
\exp\left[i\left(\qb\pb (t)-{1\over 2}q^2\right)t/m\right]
\end{align}
\end{widetext}
Strictly speaking, one should replace $\{\rb_n\}$ in the arguments of $G_x$ in the first and second terms in the right-hand side of Eq.~(\ref{eq:QKE_Wigner})  with $\{\rb_n+ \delta_{nn'}\qb (t-t')/m\}$ and $\{\rb_n- \delta_{nn'}\qb (t-t')/m\}$, respectively. For the typical values of $q\lesssim (mk_BT)^{1/2}$ and $t-t'\lesssim 1/k_BT$, this would correspond to shifting the electron coordinate by $\sim\lambdabar$, which is the uncertainty of the coordinate; in the considered regime  the corresponding change of $G_x$ should be disregarded.  

The function $G_x$ can be assumed real: the structure of Eq.~(\ref{eq:QKE_Wigner}) shows that $G_x(t) $ is real if $G_x(0)$ is real, which is indeed the case, see Eq.~(\ref{eq:G_x_general_form}) below. Therefore we are interested in the real part of the integrals over $t_1$. For the characteristic $t\gg 1/k_BT$  we have
\begin{align}
\label{eq:conservation_law}
 &{\rm Re}\,\int_0^t dt_1\xi(t-t_1); \qb,\pb)\,e^{\pm i\omega_{\qb\,\alpha}(t-t_1)}= \xi_{\pm}(\qb,\pb;\alpha),\nonumber\\
 &\xi_{\pm}(\qb,\pb;\alpha)=\pi\delta\left[\frac{1}{m}\left(\qb\pb-{1\over 2}q^2\right)\pm\omega_{\qb\,\alpha}\right],
 \end{align}
which is nothing but the energy conservation law: the change of the kinetic energy of an electron is equal to the energy of the absorbed/emitted helium vibration. Interestingly, polaronic effects drop out from the equation for the many-electron density matrix of the form (\ref{eq:G_x_general_form}). Formally, this is because we consider diagonal matrix elements on the eigenfunctions of the many-electron Hamiltonian.

\subsubsection{The many-electron relaxation time}

We now use the explicit form of the operator $\hat{\cal G}_x$, Eq.~(\ref{eq:g_x_operator}). The corresponding form of the Wigner transform is 
\begin{align}
\label{eq:G_x_general_form}
G_x\left(t;\{\pb_n\},\{\rb_n\}\right) =&  Z_{ee}^{-1}g(t)P_x \exp\left[-\beta H_{ee}\left(\{\pb_n\},\{\rb_n\}\right)\right],\nonumber\\
&g(0) = 1.
\end{align}
From Eqs.~(\ref{eq:P_correlator_Wigner}) and (\ref{eq:G_x_general_form}), the correlation function of the total electron momentum is simply expressed in terms of the function $g(t)$,
\begin{align}
\label{eq:correlator_explicit}
\langle \hat P_x(t)\hat P_x(0)\rangle = Nmk_BTg(t),
\end{align}
where $N = n_SS$ is the total number of the electrons. 

As seen from Eq.~(\ref{eq:QKE_Wigner}), the value of $G_x\left(t;\{\pb_n\},\{\rb_n\}\right)$ is coupled to the values of this function with the momentum of one of the electrons (and thus of the whole many-electron system) incremented by $-\qb$ and the energy changed by $\omega_{\qb\,\alpha}$; these values are then summed over $\qb, \alpha$. Ssubstituting Eq.~(\ref{eq:G_x_general_form}) into Eq.~(\ref{eq:QKE_Wigner}) we obtain
\begin{align}
\label{eq:q_term_explicit}
&\frac{\partial G_x}{\partial t} = -2Z_{ee}^{-1}g(t)e^{-\beta H_{ee}\left(\{\pb_n\},\{\rb_n\}\right)}\sum_{\qb,\alpha}
q_x\left|V_{\qb\,\alpha}\right|^2\nonumber\\
&\sum_{n'}\left[
\xi_{+}(\qb,\,\pb_{n'};\alpha)\bar n_{\qb\,\alpha} +
\xi_{-}(\qb,\,\pb_{n'};\alpha)(\bar n_{\qb\,\alpha}+1)\right]
\end{align}
Here we have used that, with the account taken of the energy conservation condition (\ref{eq:conservation_law}), $\xi_{\pm}(\qb,\pb_{n'};\alpha)H_{ee}\left(\{\pb_n-\qb\delta_{nn'}\},\{\rb_n\}\right) = \xi_{\pm}(\qb,\pb_{n'};\alpha) [H_{ee}\left(\{\pb_n\},\{\rb_n\}\right) \pm \omega_{\qb\,\alpha}]$.  

To evaluate the right-hand side of Eq.~(\ref{eq:q_term_explicit}) we will use an approach, that differs from that used in Ref.~\onlinecite{Dykman1997} for the case of elastic scattering. First we note that, at first glance, in the many-electron system summing over $n'$ in Eq.~(\ref{eq:q_term_explicit}) should be equivalent to averaging over the electron states for the electron system in thermal equilibrium, i.~e. to integrating $\xi_{\pm}(\qb,\pb_{n'};\alpha)$ over $\pb_{n'}$ with the weight $\propto \exp(-\beta \pb_{n'}^2/2m)$. However, the electron system has a total momentum $P_x$ along the $x$-axis. This momentum corresponds to the electrons moving along the $x$-axis with velocity $P_x/Nm$, which is the same for all electrons (one can think of watching the electron system from a frame that moves with a velocity $-P_x/Nm$). Therefore the distribution over $\pb_{n'}$ should be centered at $P_x/Nm$. Since $N\gg 1$, this means that the Boltzmann factor should be modified to 
$\exp(-\beta \pb_{n'}^2/2m)[1+\beta (p_{n'})_x P_x/Nm]$.

We also note that $|V_{\qb\,\alpha}|^2$ and $\omega_{\qb\,\alpha}$ are independent of the direction of $\qb$. The integral that describes the averaging over $\pb_{n'}$ for $P_x=0$ 
\[I_0(\qb)=\int d\pb_{n'}\xi_{\pm}(\qb,\pb_{n'};\alpha)\exp(-\beta \pb_{n'}^2/2m)\]
is independent of the direction of $\qb$. Therefore when it is multiplied by $q_x$ to calculate the right-hand side of Eq.~(\ref{eq:q_term_explicit}) and then integrated over the directions of $\qb$, the result is zero. On the other hand, for the term $\propto P_x$ the integral multiplied by $q_x$ has the form
\begin{align}
\label{eq:auxiliary_integral}
&I_\pm(\qb)=\int d\pb_{n'}\xi_{\pm}(\qb,\pb_{n'};\alpha)q_x(p_{n'})_x\exp(-\beta \pb_{n'}^2/2m)\nonumber\\
&=\frac{1}{2}\int d\pb_{n'}\xi_{\pm}(\qb,\pb_{n'};\alpha)\qb\pb_{n'}\exp(-\beta \pb_{n'}^2/2m)
\end{align}
It gives a nonzero contribution when integrated over the directions of $\qb$ (keeping in mind that $|V_{\qb\,\alpha}|^2$ and $\omega_{\qb\,\alpha}$ are independent of the direction of $\qb$, we have symmetrized $q_x(p_{n'})_x\to \qb\pb_{n'}/2$). A straightforward calculation shows that 
\begin{align}
\label{eq:I_pm}
I_\pm (\qb) = &\frac{1}{4}\left(q^2 \mp 2m\omega_{\qb\,\alpha}\right)(2\pi^3m^3/\beta q^2)^{1/2}\nonumber\\
&\times\exp\left[-\beta\left(\frac{q^2}{8m} \mp\frac{1}{2}\omega_{\qb\,\alpha}+\frac{m\omega_{\qb\,\alpha}^2}{2q^2}\right)\right]
\end{align}

Using this expression (multiplied by $\beta P_x/Nm$), we see that the right-hand side of Eq.~(\ref{eq:q_term_explicit}) takes the form 
$\tau^{-1}G_x(t;\{\pb_n\},\{\rb_n\})$ where
\begin{align}
\label{eq:tau_many_electron}
\tau^{-1} = &{1\over 2mk_BT}\sum_{\bf q,\alpha}q^2
\left|V_{\qb\,\alpha}\right|^2\left[\langle \xi_+(\qb,\pb;\alpha)\rangle \bar n_{\qb\,\alpha} \right.\nonumber\\
&+\left.\langle \xi_-(\qb,\pb;\alpha)\rangle (\bar n_{\qb\,\alpha}+1)\right]
\end{align}
where the averaging of $\xi_{\pm}$ means integration over $\pb$ with the weight $(2\pi mk_BT)^{-1}\exp(-\pb^2/2mk_BT)$, so that
\[\langle\xi_\pm(\qb,\pb;\alpha)\rangle =\left(\frac{\pi m \beta}{2q^2}\right)^{1/2}\exp\left[-\frac{\beta}{2m} \left(\frac{1}{2}q\mp \frac{m\omega_{\qb\,\alpha}}{q}\right)^2\right] \]  
The time $\tau^{-1}$ gives the static many-electron conductivity, 
\begin{align}
\label{eq:conductivity_explicit}
\sigma_{xx} = e^2n_s/m\tau^{-1}.
\end{align} 

The above analysis can be immediately extended to the electron transport in a classical magnetic field normal to the electron layer, where the cyclotron frequency $\omega_c\ll k_BT/\hbar$. In this case in the expression (\ref{eq:short_time_expansion}) for $\hat \rb_n(t')$  one should take into account the fluctuational field that drives an electron ${\bf E}_{\rm fl}$ because of the electron density fluctuations and also the cyclotron motion, cf. \cite{Dykman1997}. The relaxation rate due to inelastic scattering is weakly affected by the magnetic field for $|e E_{\rm fl}|\lambdabar \gg \hbar\omega_c$, since the discreteness of the Landau levels is smeared out by the fluctuational field. However, in the opposite case, $|e E_{\rm fl}|\lambdabar \ll \hbar\omega_c$, the discreteness of the Landau level will modify the rate of inelastic scattering. The analysis of this behavior is beyond the scope of this paper.  

\subsection{An alternative derivation}

The many-electron conductivity can be derived also using the fact that, if the electron system moves in an electric field with a velocity $\vb$,  this means that the force from the electric field is balanced by the force from the scattering of electrons  off helium vibrations. The latter force is given by the change of the total electron momentum  per unit time due to the scattering $d\Pb(\vb)/dt$. If the electric field is weak and the velocity is proportional to the field, it is easy to see that the conductivity is 
\begin{align}
\label{eq:conductivity_shortcut}
\sigma_{xx} = -e^2 n_s Nv_x(dP_x/dt)^{-1},
\end{align}
where $v_x$ and $dP_x/dt$ are the $x$ components of the corresponding vectors and the $x$ axis is chosen to point along the electric field. 

The problem of finding the many-electron conductivity is then reduced to calculating $d\Pb/dt$ for a given $\vb$ and for a given coupling to the helium excitations. This approach was developed for electrons on helium earlier for several limiting cases \cite{Dykman1979a,Buntar'1987,Vilk1989,Dykman1997b}. Here we formulate it in a general case, but assuming that no magnetic field is applied to the electron system.

The total force on the electron system due to the coupling (\ref{eq:coupling_SM}) is
\begin{align}
\label{eq:momentum_change_exact}
&\frac{d\hat\Pb}{dt} =  -i\sum_{\qb,\alpha}\qb V_{\qb\,\alpha}\hat\rho_\qb (\hat a_{\qb\,\alpha}+ \hat a_{-\qb\,\alpha}^\dagger),
\nonumber\\
&\hat \rho_\qb =  \sum_n \exp(i\qb\hat\rb_n)
\end{align}
We will calculate the expectation value of this force in the interaction representation, cf. Eq.~(\ref{eq:operator_QKE}). In this representation, if the electron system moves with velocity $\vb$, the electron density operator becomes 
\begin{align}
\label{eq:density_operator}
\hat\rho_\qb(t|\vb) = \hat\rho_\qb(t)e^{i\qb \vb t}, \quad \hat\rho_\qb(t) = e^{i\hat H_{ee} t}\hat\rho_\qb e^{-i\hat H_{ee} t}
\end{align}

To the lowest order of the perturbation theory, the expectation value of the force (\ref{eq:momentum_change_exact}) can be obtained by finding the linear response of the density matrix of the electron-vibrational system to the coupling $H_i$. The result for the real part of the force is
\begin{align}
\label{eq:momentum_change_mean}
&\left\langle\frac{d\hat \Pb}{dt}\right\rangle=-{\rm Re}\,\int_0^\infty dt\sum_{\qb,\alpha} \qb |V_{\qb\,\alpha}|^2
(1-e^{-\beta \qb\vb})\nonumber\\
&\times\langle\hat \rho_\qb(t)\hat\rho_{-\qb}(0)\rangle \phi_{\qb\,\alpha}(t) 
\end{align}
Using the single-site approximation to calculate the electron density correlator, $\langle\hat \rho_\qb(')\hat\rho_{-\qb}(0)\rangle \approx N\langle \exp[i\qb\hat \rb_n(t)]\exp[-i\qb\hat \rb_n(0)]\rangle$ and taking into account that the system is isotropic, so that when calculating the linear in $\vb$ term, in the sum over $\qb$ one can replace $\qb(\qb\vb)\to \vb (q^2/2)$, we obtain from Eqs.~(\ref{eq:conductivity_shortcut}) and (\ref{eq:momentum_change_mean}) the same expression for the conductivity as Eq.~(\ref{eq:conductivity_explicit}). 

We should note that the approximation used to obtain Eq.~(\ref{eq:momentum_change_mean}) is essentially equivalent to the approximation used to obtain the quantum kinetic equation (\ref{eq:QKE_operator_explicit}). The derivation used to obtain Eq.~(\ref{eq:tau_many_electron}), although longer, shows more clearly the approximations involved.

%

\end{document}